\documentclass[12pt,preprint]{aastex} 
 
\shorttitle{Broadband spectra of GRO~J1655-40} 
\shortauthors{Migliari et al.} 
 
\begin{document} 
 
 

\title{Tracing the jet contribution to the mid-IR over the 2005 outburst of GRO J1655-40 via broadband spectral modeling}

 
 
\author{S. Migliari\altaffilmark{1}, J.A. Tomsick\altaffilmark{1,2}, S. Markoff\altaffilmark{3}, E. Kalemci\altaffilmark{4}, C.D. Bailyn\altaffilmark{5}, M. Buxton\altaffilmark{6},  S. Corbel\altaffilmark{7}, R.P. Fender\altaffilmark{8}, P. Kaaret\altaffilmark{9}} 
 

 

\altaffiltext{1}{Center for Astrophysics and Space Sciences, 
University of California San Diego, 9500 Gilman Dr., La Jolla, CA
92093-0424, USA.} \altaffiltext{2}{Space Sciences
Laboratory/University of California Berkeley, 7 Gauss Way, Berkeley,
CA 94720-7450, USA.}\altaffiltext{3}{ Astronomical Institute `Anton
Pannekoek', University of Amsterdam, Kruislaan 403, 1098 SJ,
Amsterdam, The Netherlands.}\altaffiltext{4}{Sabanci University,
Orhanli - Tuzla, Istanbul, 34956, Turkey}\altaffiltext{5}{Department
of Astronomy, Yale University, P.O. Box 208101, New Haven, CT 06520,
USA.}\altaffiltext{6}{Department of Astronomy, Columbia University,
550 West 120th Street, New York, NY 10027, USA.}
\altaffiltext{7}{AIM - Unit\'e Mixte de Recherche CEA - CNRS -
Universit\'e Paris VII - UMR 7158, CEA-Saclay, Service
d'Astrophysique, 91191 Gif-sur-Yvette Cedex,
France.}\altaffiltext{8}{School of Physics and Astronomy, University
of Southampton, Hampshire SO17 1BJ, United
Kingdom.}\altaffiltext{9}{Department of Physics and Astronomy,
University of Iowa, 751 Van Allen Hall, Iowa City, IA 52242, USA.}

 
\begin{abstract} 
We present new results from a multi-wavelength
(radio/infrared/optical/X-ray) study of the black hole X-ray binary
GRO J1655-40 during its 2005 outburst. We detected, for the first
time, mid-infrared emission at 24~$\mu$m from the compact jet of a
black hole X-ray binary during its hard state, when the source shows
emission from a radio compact jet as well as a strong non-thermal hard
X-ray component. These detections strongly constrain the optically
thick part of the synchrotron spectrum of the compact jet, which is
consistent with being flat over four orders of magnitude in frequency.
Moreover, using this unprecedented coverage, and especially thanks to
the new Spitzer observations, we can test broadband disk and jet
models during the hard state.  Two of the hard state broadband spectra
are reasonably well fitted using a jet model with parameters overall
similar to those previously found for Cyg~X-1 and
GX~339-4. Differences are also present; most notably, the jet power in
GRO~J1655-40 appears to be a factor of at least $\sim3-5$ higher
(depending on the distance) than that of Cyg~X-1 and GX~339-4 at
comparable disk luminosities.  Furthermore, a few discrepancies
between the model and the data, previously not found for the other two
black hole systems for which there was no mid-IR/IR and optical
coverage, are evident, and will help to constrain and refine
theoretical models.
 
\end{abstract}  
 
 
 

 
\section{Introduction} 
 
Galactic black hole (BH) X-ray binaries (XRB) spend most of their time
in quiescence (a notable exception is Cyg X-1; e.g., Wilms et
al. 2006), but occasionally show transient outbursts resulting in an
increase in luminosity of many orders of magnitude at all
wavelengths. These outbursts are explained as the result of disk
instabilities, possibly due to a dramatic increase of mass accretion
rate.  The outbursts of BHs have been extensively monitored at all
wavelengths: In the radio band, infrared (IR), optical, X-rays, and up
to $\gamma$-rays.  Each observing band provides distinct windows on
the radiative processes related to the different components in the
binary systems. In X-rays, we observe the regions of the systems close
to the compact object: Inner disk, Comptonizing corona (e.g., Zdziarski
et al. 1998; Nowak et al. 1999) and/or external Compton, and
synchrotron self-Compton from the base of a jet (see e.g., Markoff,
Nowak \& Wilms 2005). In the radio band, we observe synchrotron
radiation from a relativistic jet (e.g., Fender 2006 for a review). In
the optical/IR band, three components may overlap: The outer and
irradiated disk, the companion star, and the jet (e.g., Russell et
al. 2006).
 
In X-rays, the different stages of an outburst can be described in
terms of transitions between X-ray states. The definitions of the
X-ray states are based on X-ray spectral and temporal behavior, but
their details are still under debate (e.g., Homan \& Belloni 2005;
Homan et al. 2005; McClintock \& Remillard 2006).  
In this work, we will follow the nomenclature in Remillard \&
McClintock (2006), in particular: 1) {\it Thermal} (or soft) state, when the disk flux fraction
in the 2-20~keV energy spectrum is above 75\%, the quasi-periodic
oscillation (QPO) in the power density spectrum is absent or weak and
the 2-10~keV power continuum has an integrated rms noise level
$<$7.5\%, 2) {\it Hard} state, when the fraction of flux of the
power-law component is $>$80\%, the power-law spectral index is
$1.4<\Gamma<2.1$, and the power density spectrum rms is $>$10\%.
 
These X-ray spectral states are also associated with a specific radio
(jet) behavior (see Fender 2006 for a review).  During the hard X-ray
state (i.e., quiescence to rise of the outburst), the accretion rate is
usually below 10\% the Eddington limit and the X-ray spectrum is
dominated by non-thermal power-law emission.  The model to explain
this non-thermal radiation is an area of controversy; the two
alternate scenarios currently in consideration are a Comptonizing
corona of hot electrons above an accretion disk, and
Comptonizing electrons in the base of a jet, with a contribution also
from synchrotron emission.
A steady `compact jet' is observed during this spectral state (see
Fender 2006 for a recent review).  The compact jet is characterized by
an optically thick ($\alpha\gtrsim0$, where
$S_{\nu}\propto\nu^{\alpha}$ and $S_{\nu}$ is the radio flux density
at a frequency $\nu$) synchrotron radio spectrum, and it has been
identified spectrally in many sources and spatially resolved in two BH
XRBs: Cyg X-1 (Stirling et al. 2001) and GRS~1915+105 (Dhawan, Mirabel
\& Rodr\'\i{}guez 2000).  The structure of the disc that can lead to such
jet is still being debated, although magneto-hydrodynamical
simulations seem to suggest a geometrically thick disc (Meier 2001).
During the thermal X-ray state, the thermal component can be modeled
with an optically thick, geometrically thin accretion disc (Shakura \&
Sunyaev 1973), the radio emission is quenched, likely due to a
physical suppression of the compact jet (Fender et al. 1999; Corbel et
al. 2004; Fender et al. 2004).  The hard-to-soft state transition is
likely associated with optically thin radio flares, a signature of the
ejection of transient jets (e.g., Gallo et al. 2005).

\subsection{GRO~J1655-40} 
 
The BH XRB GRO J1655-40 was the second superluminal jet source
discovered in our Galaxy (Tingay et al. 1995; Hjellming \& Rupen
1995). The mass of the compact object has been dynamically estimated
to be $M=6.3\pm0.5 M_{\odot}$, and from the optical photometry also an
inclination of the binary of $70^{\circ} .2\pm 1^{\circ}.9$ has been
derived (Greene, Bailyn \& Orosz 2001). From Very Long Baseline
Interferometry (VLBI) observations of the transient radio jets of GRO
J1655-40, Hjellming \& Rupen (1994) derived, using a distance of 3.2
kpc (in agreement with previous estimates: McKay \& Kesteven 1994,
Tingay et al. 1995), a jet axis inclination of $\sim 85^{\circ}$ to
the line of sight, with a possible precession of the jet around the
axis of $\sim2^{\circ}$. Foellmi et al. (2006), based on the estimated
optical absorption towards GRO J1655-40, have recently placed an upper
limit on the distance to the source of $\sim1.7$~kpc. With this new
distance, the transient jets that were previously observed would no
longer be superluminal. Also, using the lower distance of 1.7 kpc, the
inclination of the jet axis as derived by the VLBI observations would
be a few degrees lower. In this paper we will use the distance of
1.7~kpc for our calculations and fits. As a caveat, note that the
distance is still under debate. Another work, still in preparation,
argues that a distance greater than 3 kpc is required to
explain the ellipsoidal variations observed in the optical and
near infrared (Bailyn et al. 2007, in prep.).

After seven years of quiescence, GRO J1655-40 entered a new outburst 
on February 2005, when the source showed an increase in the X-ray flux 
(Markwardt \& Swank 2005), optical and near-IR magnitude (Torres et 
al. 2005; Buxton, Bailyn \& Maitra 2005) and renewed radio activity 
(Rupen, Dhawan \& Mioduszewski 2005a). The outburst lasted about eight 
months and has been extensively followed, when possible on a daily 
basis, at all wavelengths. In March, a state transition occurred as 
GRO~J1655-40 entered a thermal state and the radio counterpart faded 
(Homan 2005; Rupen, Dhawan \& Mioduszewski 2005b). In May, the source 
entered a highly variable, high X-ray luminosity state (Homan et 
al. 2005a) coupled with renewed radio emission (Rupen, Dhawan \& 
Mioduszewski 2005c). GRO~J1655-40 then entered a soft state, with no 
radio detection and returned to a hard state on September 23 (Homan 
et al. 2005b). The source returned to radio activity on September 21 
(Brocksopp et al. 2005). 
 
In this work, we present new results from a multi-wavelength
(radio/infrared/optical/X-ray) campaign of GRO J1655-40 during its
2005 outburst. We study the broadband spectral energy distribution
during the different stages of the outburst, with particular emphasis
on the important new simultaneous observations in mid-infrared by
Spitzer, which also allows for new constraints on the jet scenario.
We provide an overview of the radio, mid-/near-IR, optical, soft/hard
X-ray observations and data analysis in \S~\ref{observations}; 
we discuss the evolution of the spectra during the outburst in 
\S~\ref{evolution}, the detection of the  mid-IR emission from the compact jet in
\S~\ref{IR-jet},  and the results of the fit of the broadband spectra
in the context of a jet model and the discussion in
\S~\ref{jet-model}.
 
\section{Observations}\label{observations} 
 
We have observed GRO~J1655-40 with Multiband Imaging Photometer for
Spitzer (MIPS) during its outburst that started in 2005, following the
different stages from the rise of the outburst until quiescence in
2006: 1) in hard state during the rise on 2005 March 10, 2) in a
thermal state after the first X-ray flux peak on 2005 April 6, 3) in
a thermal state after the second and brightest X-ray flux peak on
2005 August 28, 4) during the decay of the outburst, immediately
after the BH returns in the hard state on 2005 September 23 and,
finally, 5) after the outburst ended, during quiescence on 2006,
April~1. The arrows on top on Fig.~1, lower panel, show when these
observations have been performed with respect to the 2-12~keV light
curve of the All Sky Monitor (ASM) onboard the Rossi-X-ray Timing
Explorer (RXTE).  Since the 2005 outburst started, GRO~J1655-40 has
been monitored daily in X-rays with pointed RXTE observations (Jeroen
Homan and coworkers\footnote{http://tahti.mit.edu/opensource/1655/}),
and INTErnational Gamma-Ray Astrophysics Laboratory (INTEGRAL;
Integral Galactic bulge
group\footnote{http://isdcul3.unige.ch/Science/BULGE/SOURCES/GRO\_J1655-40/GRO\_J1655-40.html}),
in optical/near-IR with the Small and Medium Aperture Research
Telescope System (SMARTS; Michelle Buxton and Charles
Bailyn\footnote{http://www.astro.yale.edu/buxton/smarts/light\_curves/oir\_gro.jpg})
and with a good coverage also in the radio band with the Very Large
Array (VLA; Michael Rupen and
coworkers\footnote{http://www.aoc.nrao.edu/\%7Emrupen/XRT/GRJ1655-40/grj1655-40.shtml};
see also Shaposhnikov et al. 2006).  The four Spitzer/MIPS
observations during the outburst (see Fig.~1, top panels) were all
simultaneous with pointed RXTE, SMARTS and radio (either with the VLA
or the Australia Telescope Compact Array) observations, allowing us to
study the evolution of the complete broadband spectrum of the BH
during the different stages. GRO~J1655-40 has also been observed with
the Spitzer/Infrared Array Camera (IRAC) in the hard state on
September 29, simultaneously with RXTE, SMARTS and
quasi-simultaneously (on 2005 October 2) with VLA observations.  The
Spitzer/MIPS observation in quiescence, on 2006 April 1, had no
coverage at other wavelength, except for the 2-12~keV observations of
the RXTE/ASM. The logs of the RXTE and Spitzer observations are shown
in Table~1.
 
\subsection{Infrared and mid-infrared: Spitzer IRAC and MIPS} 
 
We have processed the Basic Calibrated Data of the MIPS observations
at 24~$\mu$m and IRAC observations at 3.6, 4.5, 5.8 and 8~$\mu$m using
the software {\tt mopex} (Makovoz \& Marleau 2005).  We created
mosaics from the 70 and 59 frames per band obtained in the MIPS and
IRAC observations, respectively.  GRO J1655-40 is a few arcseconds
south of an extended mid-IR emitting source, very bright at 24 $\mu$m
(see Fig.~1, top panels), which enhanced the background in the region
of the BH and increases the uncertainties in its flux estimates. GRO
J1655-40 is observed to vary significantly over our five MIPS
observations, and the mid-IR emission appears to be off when the
source is in its quiescent state, on 2006 April 1.  We extracted the
flux density in a circular region centered at the optical coordinates
of GRO J1655-40 and with a radius of 10 arcsec in the quiescent state
observation, and used this flux (540~$\mu$Jy) as background for the
estimate of the flux density of the source in the other MIPS
observations. We extracted the flux density of GRO J1655-40 in the
other five MIPS observations and in the IRAC observation using
aperture photometry, with a 10 arcsec radius circle.  For each
observation, we created the point-response functions (PRFs) with {\tt
prf\_estimate} and calculated the aperture corrections using the
extracted PRF.  We have corrected for interstellar extinction using
$A_{\rm v}=3.72$, derived from Greene, Bailyn \& Orosz (2001) and
following the standard optical-to-IR interstellar extinction law
(e.g., Rieke \& Lebofsky 1985; Cardelli, Clayton \& Mathis 1989). Note
that, following Foellmi et al. (2006) instead, we would have obtained
a slightly lower value of $A_{\rm v}=3.53$.  We added $5\%$ and $10\%$
systematic errors on the estimate of the flux densities in the IRAC
and 24 $\mu$m MIPS observations, respectively, to take into account
the uncertainties in the photometric calibration (see Reach et
al. 2005). The flux densities of the Spitzer observations are shown in
Table~2.
 
\subsection{Optical/near-infrared: SMARTS}

Optical and IR monitoring of GRO J1655-40 was carried out throughout
the 2005 outburst using the SMARTS consortium telescopes.  Starting
2005 February 21, observations were carried out each clear night with
the 1.3m telescope at CTIO, and the ANDICAM instrument.  The ANDICAM
(Depoy et al. 2003) is a dual-channel imager containing an optical CCD
and an IR array, so simultaneous observations can be obtained from one
optical (BVRI) and one IR (JHK) bandpasses.  In the case of the
outburst of GRO J1655-40, nightly observations were obtained in B, V,
I, J and K (Buxton, Bailyn \& Maitra 2005; Buxton \& Bailyn 2005).
The full SMARTS light curve will be presented elsewhere.

Standard flatfielding and sky subtraction procedures were applied to
each night's data, and the internal dithers in the IR were combined as
described in Buxton \& Bailyn (2004).  Differential photometry was
carried out each night with a set of reference stars in the field.
Intercomparisons between reference stars of similar brightness to the
source suggest a precision of $<0.02$ magnitudes in BVIJ, and $\approx
0.03$ magnitudes in K.  Calibrations to the standard optical magnitude
system were carried out using Landolt standards (Landolt 1992),
several of which were observed on each photometric night, together
with extinction corrections calculated from these standards over the
course of the entire 2005 observing season.  IR magnitudes were placed
on the 2MASS system using 2MASS stars present in the field of view of
GRO J1655-40.  We estimate the accuracy of our standard field
calibration to be better than 0.05 magnitudes in all bands. We show
the apparent magnitudes, not yet de-reddened, in Table~3.

\subsection{X-rays: RXTE} 
 
\subsubsection{X-ray spectral analysis} 
 
We have analyzed the RXTE pointed observations performed
simultaneously with our Spitzer observations. We have used the PCA
{\tt Standard2} data of the proportional counter unit 2 (PCU 2), which
was on in all the observations, to produce the hardness-intensity
diagram (HID) shown in Fig.~\ref{fig:hid}.  The hard color is defined
as the count rate ratio (9.4-18.5)~keV/(2.5-6.1)~keV.  For the energy
spectral analysis, we have used PCA {\tt Standard2} of all the PCUs
available, and HEXTE {\tt Standard Mode} cluster A and B data. For the
PCA data, we have subtracted the background estimated using {\tt
pcabackest} v.3.0, produced the detector response matrix with {\tt
pcarsp} v.10.1, and analyzed the energy spectra in the range
3--25~keV. A systematic error of 0.5\% was added to account for
uncertainties in the calibration. For the HEXTE data, we corrected for
deadtime, subtracted the background, extracted the response matrix
using FTOOLS v.6.1.2.  We have analyzed the HEXTE spectra between 20
and $200$~keV.  The 3-200 keV spectra are well fitted using a
multicolor disk black-body, a power-law, with a cutoff for the March
10 and April 6 spectra, a smeared edge around 7-9~keV (we constrained
the width of the edge to be $<15$~keV) and, for the April 6
observation, a Gaussian emission line around 6.2 keV is required,
possibly a red-shifted Iron line, although the energy is still
marginally consistent with neutral Iron line centered at 6.4~keV.  We
also accounted for photoelectric absorption from interstellar
material.  The inner temperature of the disk is particularly low for
the three hard state observations, and we fixed it to 0.5~keV because
it cannot be well-constrained.  Also, we fixed the equivalent hydrogen
column density to $N_{\rm H}=8\times10^{21}$~cm$^{-2}$, a value
comparable to those measured in previous observations of GRO~J1655-40
(e.g., Tomsick et al. 1999). We show the best-fit parameters of each of
the 3-200 keV spectra in Table~4.

\subsubsection{X-ray temporal analysis} 
 
For each observation, we compute the power density spectra from the
PCA data using IDL programs developed at the University of T\"ubingen
(Pottschmidt 2002). The power density spectrum is normalized as
described in Miyamoto \& Kitamoto (1989) and corrected for the
dead-time effects according to Zhang et al. (1995). Using 256 second
time segments, we investigate the low frequency quasi-periodic
oscillations (QPOs) and the timing properties of the continuum up to
100~Hz. We fit all the power density spectra with broad and narrow
Lorentzians (Fig.~3) with our standard timing analysis techniques
(e.g., Kalemci et al. 2005; see Belloni, Psaltis \& van der Klis 2000;
Nowak 2000; Pottschmidt et al. 2003).  The rms amplitudes are
calculated over the whole frequency range of the power density
spectrum, calculated from zero to infinity from the fitted
Lorentzians, integrated over 2-15~keV. 
Although the aperiodic X-ray variability features are still poorly
understood in detail, they are thought to be related to physical time scales in
the accretion disk (see, e.g., van der Klis 2006 for a review).
The multi-Lorentzian model description of the power spectra, although
not necessarily physically motivated, makes it  possible to identify
the different components in the power density spectrum and follow
their variations also in relation with other observational parameters
(see Belloni, Psaltis \& van der Klis 2000). These characteristics
make the power density spectra a powerful (complementary) tool to
classify the different observational states of the X-ray binaries. In
this work, we will study the power density spectra only for
classification purposes, following Remillard \& McClintock (2006) and
Homan \& Belloni (2005) (see also van der Klis 2006). The power
density spectra on April 6 and August 28 are well fitted with two
broad Lorentzians, no QPOs are present and the total rms is $\lesssim
5\%$, which is typical of observations in a thermal state. More than
two Lorentzians are needed to fit the more complex power density
spectra of the March 10, September 24 and September 29
observations. At least three broad Lorentzians and one QPO component
(plus the first harmonic of the QPO in the case of the March 10
observation) are necessary. The presence of a 0.1-10 Hz QPO and the
total rms $\gtrsim 25\%$ indicate that the observations are in or very
close to a hard state. We show the power density spectra of the five
observations with the multi-Lorentzian fitting components in Fig.~3.

\subsection{Radio: VLA} 
 
In this work, we used the VLA radio flux densities at 5 and 8.5~GHz
from Shaposhnikov et al. (2006) for the 2005 March 10 observation, at
5~GHz from Brocksopp et al. (2005) for the 2005 September 22
observation, and at 5~GHz from Michael Rupen and collaborators
webpage\footnote{www.aoc.nrao.edu/$\sim$mrupen/XRT/GRJ1655-40/grj1655-40.shtml}
for the observations on 2005 April 6, August 28, and October 2. The
radio-to-X-rays spectral energy distributions (SEDs) of the 5
observations of GRO~J1655-40 are shown in Fig.~4.

\section{Results and Discussion}

\subsection{The Outburst Evolution}\label{evolution}

We follow the evolution of our six observations during the outburst
using the X-ray light curve (Fig.~1), the HID (Fig.~2), the power
density spectra (Fig.~3) and the SEDs (Fig.~4). We inspected the PCA
light curves with 16 second time resolution for the observations taken
on March 10, April 6, August 28, September 24 and September 29. We do
not see any long-term trends in the count rate over the duration of
the observations (on a time scale of hours), or any X-ray dips or
flares with amplitudes greater than $\sim15\%$.

1) Based on the X-ray definition in Remillard \& McClintock (2006), on
2005 March 10, the source is in the {\it hard state}.  The power
density spectrum shows a high rms of $\sim34$\% and broad features as
well as a narrow QPO around 2~Hz. The X-ray energy spectra show a disk
flux of $\sim5\%$ the total 2-20~keV flux, also consistent with their
definition of hard state.  However, the X-ray flux already started its
abrupt rise towards the first peak of the outburst.  The position on
the HID, if compared to those of September 24 and 29 which are also in
the hard state (see below; see also Homan's `open source'
page\footnote{http://tahti.mit.edu/opensource/1655/} for a comparison
with other GRO~J1655-40 observations), suggests that the source was
leaving the hard state. Indeed, if we follow the nomenclature of Homan
\& Belloni (2005) instead, which is based on the spectral index of the
X-ray power law, and the QPO and integrated rms strength in the power
density spectra, we would identify the March 10 as a {\it hard
intermediate state} (HIMS) observation (see also Shaposhnikov et
al. 2007).  The radio emission is significantly detected in two bands
(5~GHz and 8.6~GHz), with a spectral index of $\alpha=-0.36\pm0.34$;
this spectral index is consistent with either a compact jet, as
typically observed in hard state observations, or optically thin
synchrotron emitting jet, possibly indicating that the outer part of
the jet was decoupled from the system and the source had already left
the hard state. Given the lack of conclusive proof, we will discuss
both the hard state and HIMS classifications, where, with `hard
state', we also imply that a radio optically thick jet is present.  An
excess in the spectrum at 24~$\mu$m suggests that the jet component is
dominant also in the IR band (see \S~\ref{IR-jet} for a
discussion). In case the source is in a hard state and the radio
emission is from a compact jet, a power-law fit of the radio-to-IR
spectrum gives an almost flat spectral index of $\alpha=0.08\pm0.03$.
 
2) On 2005 April 6, the source is in a {\it thermal state}. The
X-ray light curve shows that during this observation, GRO~J1655-40 is
in a steady high flux state, in between the two outburst peaks. The
power density spectrum shows a {\em rms}$\sim5\%$, typical of a
thermal state. The power-law component in the X-ray spectrum is about
$20\%$ of the total 2-20~keV flux and the source is, accordingly, in
the upper-left, soft region of the HID pattern. The radio emission is
already quenched with a 5~GHz 3$\sigma$ upper limit of 1~mJy, and the
thermal emission dominates the energy spectrum in the X-ray, optical
and mid-IR band: Spitzer/MIPS detected the IR tail of the bright disk
at 24~$\mu$m. Note also that the hard X-ray component above
$\sim30$~keV disappears in this observation, going below the detection
threshold of HEXTE. 
 
3) On 2005 August 28, GRO~J1655-40 is in a {\it thermal state}.  The
X-ray light curve shows that the source is still at a high flux level,
but already starting its decay towards the hard state. The {\em rms}
noise in the power density spectrum is $\sim2\%$, typical of a thermal
state, as is its position on the far left of the HID. The energy
spectrum still shows a bright disk in the soft X-rays, where the disk
flux is $94\%$ of the total 2-20~keV flux.  We also see the
reappearance of the hard X-ray component above 30~keV. The radio
emission is not detected down to a $3\sigma$ upper limit of
$\sim1$~mJy and the source is only marginally detected at 24~$\mu$m.

4-5) On 2005 September 24 and 29, GRO~J1655-40 is observed during
the decay of the outburst, when it returns to the {\it hard
state}. The {\em rms} values in the power density spectra increase
significantly to $\sim25\%$ and some features, like a QPO around
0.3~Hz appears on September 24. The source reaches the bottom right
part of the HID and the X-ray spectra are dominated by a non-thermal
power-law component whose 2-20~keV flux is more than 90\% of the total
flux.  The IR emission (IRAC on September 24 and MIPS on September 29)
shows an excess due to the re-brightening of the jet. This jet
re-brightening is clearly visible in the radio band, where its flux
density at 5~GHz increases between September 24 and 29, contrary to
the X-ray flux that is still decaying in time: no radio/X-ray flux
positive correlation is present.
 
6) On 2006 April 1, the source has already returned to quiescence; no
pointed RXTE, radio and optical observations are available. The
Spitzer/MIPS observations does not detect the source at 24~$\mu$m. \\

\subsection{Mid-infrared emission from the compact jet}\label{IR-jet}

We detect mid-IR emission at 24~$\mu$m from the hard state
observations of GRO~J1655-40 on September 29 and the hard (or HIMS)
observation on March 10. The source is also detected on April 6, when
GRO~J1655-40 was in a thermal state. Thanks to the optical and near-IR
simultaneous observations, we can clearly distinguish the contribution
of the companion star in the binary system; its spectrum can be
represented by a black body peaking at a few $10^{14}$~Hz, that should
show a Rayleigh-Jeans decay at lower frequencies. The comparison of
the near-IR flux distribution with a power-law with spectral index 2
and normalized to the flux in the K band ($1.39\times 10^{14}$~Hz),
clearly shows in both the March 10 and the September 29 observation, a
deviation from a Rayleigh-Jeans spectrum in the mid-IR, with an excess
at 24~$\mu$m. Other possible contributors to the mid-IR emission are
the jet and the irradiated disk components (Cunningham 1976; Vrtilek
et al. 1990; Hynes et al. 2002; see also Russell et al. 2006 for a
discussion). The variability observed in the mid-IR rules out a
circumbinary disk origin. 

In the March 10 observation, the disk emission is significantly higher
than in the other hard state observations and the contribution of the
disk irradiation might be significant.  Mid-IR emission is also
detected during the April 6 observation, when GRO~J1655-40 was in a
thermal state. The disk component dominates the X-ray spectrum, and a
black body the optical-IR band (see Fig.~4). The jet is not detected
in the radio band with a $3\sigma$ upper limit of $1$~mJy, supporting
the evidence that in BHs the compact jet is suppressed during the
thermal state. In this case, the mid-IR emission is most likely due to
the contribution of an irradiated disk component. The 24~$\mu$m flux
of the April 6 and of the March 10 observations are comparable (see
Table~2). However, during the March 10 observation both the disk and
the black body component are much fainter, indicating, therefore, a
significant contribution in the mid-IR of another component, i.e., the
jet, which is also detected in the radio band.

In the September 29 observation, the mid-IR emission is about 50\%
higher than on March 10 (Table~2), but the disk emission is much
weaker (Table~4), strongly indicating that the compact jet, also
clearly detected in the radio band, is the dominant contributor to the
mid-IR. Indeed, a fit with a power-law model from the radio band to
the mid-IR, gives a spectral index of $\alpha=0.07\pm0.04$, consistent
with a flat optically thick synchrotron emission from a compact jet. A
deviation from a Rayleigh-Jeans spectrum is observed also on September
24, where an excess flux is already present at 8~$\mu$m. This excess,
as well as in the September 29 observation, is likely dominated by the
jet.

\subsection{Modeling the SEDs: New Constraints on Jet Models}\label{jet-model}

Different theoretical models exist to explain the emission mechanism
for BH X-ray binaries in the hard state (for a discussion, see
e.g., Tomsick, Kalemci \& Kaaret 2004 and references therein).  One of
the possibilities that we will further explore in this paper is that
the compact jet, observed to produce synchrotron emission from the
radio to at least the IR band, is also responsible for the hard X-ray
emission, which comes from the base of the jet (e.g., Markoff, Falcke
\& Fender 2001; Markoff \& Nowak 2004; Markoff, Nowak \& Wilms 2005).
Difficulties in reproducing some of the observed features seem to
disfavor the direct synchrotron emission from the base of the jet as
the `dominating' X-ray emitting mechanism in at least a few hard state
observations, as in the case of the BH 4U~1543-47 (Kalemci et
al. 2005).  On the other hand, K\"ording, Falcke \& Corbel (2006),
based on statistical analysis of the sample of BH systems in the
`fundamental plane' (Merloni, Heinz \& Di Matteo 2003; Falcke, Markoff
\& K\"ording 2004; see also the discussion in Heinz 2004), pointed out
that a synchrotron/jet scenario, discussed in the more updated
prescription including external Compton and SSC, is still in agreement
with the fundamental plane for hard state BHs. Other possible
arguments against the synchrotron-only jet model as the dominant X-ray
emission mechanisms in X-ray binaries, have been discussed in
e.g., Maccarone (2005 and references therein). In the following, we
will test the most updated version of the `jet model' which includes
also SSC and external inverse Compton. Other models exist to interpret
the observed broadband spectra, and, to date, there are no {\it
conclusive} arguments which favor one in particular.  Yuan et
al. (2005) proposed, for example, an accretion-jet model to interpret
the broadband energy spectra of the BH XTE J1118+480 in hard state
(the same source also successfully fitted with the earlier version of
the `jet model' in Markoff, Falcke \& Fender 2001), where a simple
synchrotron emitting compact jet is superimposed {\it ad hoc}, to a
`hot accretion flow' model fitting the X-ray spectrum (i.e., an outer
thin disk cohexisting with an inner Advection Dominated or
Luminous-Hot Accretion Flow). In their model, the hard X-ray emission
comes from the hot accretion flow through thermal Comptonization.
However, this model is not yet testable statistically and
self-consistently in the whole broadband spectrum. Indeed, among the
different models available to test against our new broadband SEDs, the
jet model described in Markoff, Nowak \& Wilms (2005) is the only
refined broadband model that can be tested with $\chi^2$ statistics,
from the radio band to the hard X-rays.

This jet model, in its latest prescription, has already started to be
explored by fitting the broadband energy spectra of two BH XRBs in
hard state.  Most notably, Markoff \& Nowak (2004) show that either a
Comptonizing corona or synchrotron self-Compton from the base of a
compact jet can fit the X-ray part of the energy spectra of Cyg~X-1
and GX~339-4 in hard state with the same statistical quality.  Also,
the non-thermal hard tail in the X-ray spectra of GRO~J1655-40 can be
well fit with `corona' models. The best-fit values from a simple fit
using a power law to account for the hard X-ray tails are shown in
Table~4.  One of the advantages of the jet model is that it can
interpret the whole observable broadband spectrum, from the radio band
to the highest energies, in a self-consistent manner. For Cyg~X-1 and
GX~339-4, though, the broadband spectra analyzed in Markoff, Nowak \&
Wilms (2005) relied on simultaneous radio and X-ray observations,
leaving the optical and IR portion of the spectrum uncovered. The
mid-IR, IR and optical bands are indeed critical for testing the
assumptions of the model and constraining fundamental jet model
parameters.

\subsubsection{The jet model} 
 
For a detailed discussion of the jet radiative model, we refer the
reader to e.g., Markoff \& Nowak (2004) and Markoff, Nowak \& Wilms
(2005). We recall here some fundamental assumptions and a brief
description of the model, as outlined in Markoff, Nowak \& Wilms
(2005): (1) the total power in the jet scales proportionally with the
accretion power at the inner edge of the disk, (2) the jet is
expanding freely and, at the very base, is only slightly accelerated
as a result of the pressure gradient, (3) the jet contains cold
protons that carry most of the kinetic energy while the leptons are
the dominant source of radiating energy, (4) some particles are
eventually accelerated into a power-law distribution, (5) the
power-law is maintained along the jet beyond the shock
region. Geometrically, the base of the jet is comprised of a region
with (nozzle) radius $r_{0}$, whose lower limit is the innermost
stable orbit of the disk around the black hole. The uncertainties
about the physics of jet formation are absorbed by initializing
parameters in this region, for the rest of the jet. The jet starts as
a cylindrical flow, with constant radius $r_{0}$.  After this small
nozzle region, above $\sim10$ gravitational radii ($r_{g}$), the jet
expands sideways at the sound speed for a proton/electron plasma
(i.e., $\sim0.4c$) and is only slightly accelerated by the resulting
pressure gradient. At a distance of 10-100~$r_{g}$ the particles in
the jet, that started with a quasi-thermal distribution, are
accelerated by into a power-law distribution.

To zeroth order (for a more detailed discussion of other cooling
effects already present or not yet present in the model, see Markoff,
Nowak \& Wilms 2005), the resulting jet emission spectrum is the
superposition of (1) an optically thick synchrotron spectrum coming
from the outer regions of the jet, beyond the shock region, emitting
in the radio up to likely the IR band with a flat or slightly inverted
power-law spectrum, (2) an optically thin synchrotron spectrum, still coming
from the post-shock region but emitting at frequencies above which the jet
is transparent, which emits a power-law spectrum with a negative
spectral index dependent on the electron power-law distribution, (3)
an optically thin and optically thick synchrotron emission from the
quasi-thermal distribution of particles coming from the pre-shock jet
region, and (4) external Compton from the accretion disk plus a
synchrotron self-Compton (SSC) spectrum coming from the very base of
the jet, from the nozzle region. Effects of high energy cooling are
added to this spectrum, so that the optically thin part synchrotron
spectrum decays exponentially above a certain frequency; the maximum
electron energy is calculated self-consistently for the local cooling
rate. A multicolor disk blackbody is added as an independent, fitted
spectral component, whose photons are included for upscattering in the
jet. Note that an irradiated disk component is not yet included in
this version of the code.

\subsubsection{The fits} 
 
We focus only on the hard state observations.  We fitted the energy
spectra of 2005 March 10, September 24 and September 29, using the
Interactive Spectral Interpretation System (ISIS; Houck \& Denicola
2000). This software has two main advantages: 1) we can deal easily
with broadband spectra, combining spectra with response matrices and
ASCII tables listing energy channels and flux densities, without
response matrices, 2) it can create model-independent unfolded
spectra. We refer to the work of Nowak et al. (2005) for a detailed
discussion.  As in Markoff, Nowak \& Wilms (2005), we started the fit
manually, trying to reach a $\chi^2$ such that
$\chi^{2}_{\nu}<10$. Then we use these parameters as a starting set of
parameters for the fit with ISIS. This procedure helps to avoid that
the automatic minimization in ISIS would fall in local minima. The
fitting analysis, starting from the manual fitting, is a fairly long
procedure that can take up to a week per spectrum.

The fitting model we used consists of three components, corrected for
photoelectric absorption: 1) the disk and jet models discussed above,
2) a blackbody to model the companion star, likely a F6III-F7IV
(Shahbaz et al. 1999; Israelian et al. 1999; Foellmi et al. 2006), and
3) a disk reflection component ({\tt pexriv} in ISIS) with a single
Gaussian emission line in the range 6-7~keV. The values with the
best-fit parameters of September 24 and 29 spectra are shown in
Table~5. We first fixed the physical parameters which can be
constrained by observations. We fixed the mass of the BH to
7~M$_{\odot}$ (Orosz \& Bailyn 1997; see also Greene, Bailyn \& Orosz
2000), and the distance to its most recent inferred upper limit of
1.7~kpc (Foellmi et al. 2006). In \S~1.2 we noted that the upper limit
of 1.7~kpc on the distance is still under debate and that there are
arguments that still indicate a lower limit of 3~kpc (Bailyn et
al. 2007, in prep.). A debate on the distance goes beyond the scope of
this paper, and we decided primarily to use the latest published
value.  However, since the distance enters non-linearly in many
parameters of the fitting jet model, and a simple rescaling of the
best-fit values in Table~5 is not possible for all of them, for
comparison, we report in Table~5 also the results of a fit to the
September 29 observations, with a fixed distance of 3.2~kpc.  We fixed
the inclination of the jet, at first, to $75^{\circ}$, and then we
calculated a second fit for the September 29 observation with the
inclination free as this observation requires a somewhat flatter
optically-thick synchrotron spectrum (see below for a more detailed
discussion).  An inclination of $75^{\circ}$ is consistent with the
jet axis inclination inferred from the radio lobe observations by
Hjellming \& Rupen (1999), revised with the new upper limit on the
distance to the source of 1.7~kpc. This value is also consistent with
the disk inclination of $\sim70^{\circ}$ inferred by Greene, Bailyn \&
Orosz (2002) and allowing a disk-jet misalignment of less than
$15^{\circ}$ (e.g., Maccarone 2002). We also fixed the parameters
known from previous works to fall in the same range for the other BHs
(Cyg X-1, GX~339-4; this choice is made as a starting point to explore
the parameter space, see also Markoff, Nowak \& Wilms 2005 for
discussion). The relevant fitting parameters are shown in Table~5.
 
\subsubsection{A Comparison with other BHs} 
 
In Table~5 we show the best-fit parameters of the September 24 and 29
observations (during the decay of the outburst) using the jet model,
as described above.  We show the SEDs of the observations with the
fitting model components in Figs~5, 6 and 7. The observation on March
10 during the rise of the outburst did not give a good fit
($\chi^{2}_{\nu}=8.35~(80~d.o.f.)$), and we will discuss it in more
detail in \S~3.3.4.  The jet model fit fairly well the data of
September 24, with a $\chi^{2}_{\nu}=1.72~(81~d.o.f.)$, and of
September 29, with a $\chi^{2}_{\nu}=0.90~(56~d.o.f.)$, fixing the jet
inclination to $75^\circ$. Note that the high $\chi^2$ might, at least
in part, be due to the fact that the optical and IR part of the
spectrum has been fitted with a simple black body spectrum, which was
added as an independent component to the jet model. In the September
29 fit (Fig.~\ref{sep24fit}), although the fit is good in a
statistical sense, the model somewhat underestimates the radio
emission.  The slope of the optically-thick part of the synchrotron
spectrum in the modeled jet is steeper than the slope required by the
observations.  In order to obtain a flatter radio-IR synchrotron
spectrum, we let the jet inclination be a free parameter, and the
best-fit inclination obtained is $\sim40^{\circ}$
($\chi^{2}_{\nu}=0.94~ (55~d.o.f.)$; see Fig.~7).  We would like to
stress that, at this stage, the free inclination is meant to be an
artificial modification to try to obtain a better by-eye fit in the
radio band. A jet inclination of $\sim40^{\circ}$ seems unlikely given
a disk inclination of $70^{\circ}$ (Greene, Orosz
\& Bailyn 2001) and the previous estimates of the jet inclination 
(e.g., Hjellming \& Rupen 1995), even using the smaller distance in
Foellmi et al. (2006). Furthermore, the uncertainties in modeling the
jet emission are still too large to attempt an estimate of the jet
inclination using these fits. The $\chi^2$ values for this fit are
comparable because the statistics are dominated by the X-ray and
optical part of the spectrum, but there is an improvement in the fit
in the radio band.
 
A remarkable result of these fits is that the same model that can fit
well GX~339-4 and Cyg~X-1 broadband spectra, seems not to reproduce
equally well the optically thick part of the synchrotron emission in
GRO~J1655-40, which is flatter than the model predicts. Possible ways
to make the radio-IR emission flatter would be 1) to have a less
beamed jet (which is the case emulated in the fit by the smaller jet
inclination angle), that can be due to e.g., a smaller gradient in the
regions contributing to the optically thick part of the jet and 2) to
have a more collimated jet, such as what may be expected in the case
of magnetic collimation, which is still poorly understood and not
included here.
 
As mentioned above, Markoff, Nowak \& Wilms (2005) fitted, using the
jet model, some typical hard state observations of the BH XRBs Cyg~X-1
and GX~339-4, and discussed the differences and similarities found in
the best-fit parameters. In particular, they found that 1) the total
initial power that goes into the jet $N_j$ is similar for both
sources; 2) Cyg X-1 seems to favor a more compact jet base, with the
same radius-to-height ratio ($h_0$), but with $4.4~r_g<r_0<9.1~r_g$
for Cyg X-1 against the $9.6~r_g<r_0<20.2~r_g$ of GX~339-4. The
smaller $r_0$, with a similar $N_j$, reflects the higher
X-ray-to-radio flux ratio of Cyg~X-1 with respect to GX~339-4; 3) The
temperature of the electrons in GX~339-4 is $T_{e}\sim4000$~keV, a
factor of two larger than in Cyg X-1; 4) the spectral index of the
electron distribution $p$, the fraction of accelerated electrons
$pl_f$ (which was large and then fixed to $75\%$ in this work, as
suggested in their paper), and the location of the acceleration region
$z_{acc}$ (approximately 20-30~$r_g$) are roughly the same for the two
BHs; 5) both BHs appear to have the jets close to equipartition.

Comparing the parameters of GRO~J1655-40 with those of the other two
BHs (Table~5), it is remarkable that most of the parameters are
very similar for these three different BHs. Some differences are
however present; we find that the power that GRO~J1655-40 put into
the jet is higher than that in the other two BHs, by a factor of $>3$
(the jet power is of the same order of $N_{j}$ in Table~5; see also
Appendix A2 in Markoff, Nowak \& Wilms 2005).  Also, the nozzle radius
is smaller than that of GX~339-4 and very similar to that of Cyg~X-1,
reflecting the higher X-ray/radio flux ratio. The electron temperature
is approximately the same as that found for the other two BHs, as is
the spectral index of the electron distribution. The model finds
solutions for a jet close to equipartition ($k\sim1$ in Table~5) only
in the case of September 29, but only if we let the jet inclination
free to adjust for the otherwise steeper optically thick radio-to-IR
spectrum observed. The September 24 and 29 observations with fixed
inclination to 75$^{\circ}$, find a solution for a weakly magnetically
dominated jet ($k\sim4$ and $k\sim5$, respectively), which is
different from what found for other two BH XRBs, but close to the
values found for e.g., Low-Luminosity Active Galactic Nuclei as Sgr A*
and M81 (e.g., Markoff et al. 2004). Note, however, that this
equipartition parameter is still not very well constrained and, also,
that there is no physical reason to prefer a solution with a jet in
equipartition over that of a (weakly) magnetised jet.  Using a
distance of 3.2~kpc, the jet power of GRO~J1655-40 would be about 5 times
higher than that derived in the fits of Cyg~X-1 and GX~339-4 for
comparable disk luminosities. The jet and the disk luminosities seem
to be the only two parameters that change significantly (increase) using a
distance of 3.2~kpc (see Table~5).

\subsubsection{March 10: X-rays from a corona or from the base of the jet?} 

We are not able to obtain a statistically good fit for the observation
on 2005 March 10. We fitted this broadband spectrum with the same
model components we used for the September observations, in order to
emphasize the differences between this and the other hard state
spectra. The reflection fraction parameter was left free during the
fit and reached a value of $\sim70\%$. Such a high reflection fraction
is two times the one found in the September observations (Table~5,
using a distance of 1.7~kpc) and is physically incompatible with a
mildly relativistic beamed corona or a jet model (Belodoborov 1999;
Markoff \& Nowak 2004). An edge around 6-7 keV and/or a stronger iron
line at a different peak energy might be needed to slightly improve
the spectral fitting.  However, the `hard state' jet model in the
present form seems not to be sufficient to describe statistically well
the observed broadband data. On March 10, the source is either in a
hard state or in a HIMS (according to the definitions in Remillard \&
McClintock 2006 or Homan \& Belloni 2005, respectively). The radio
emission is present and the radio spectral index is consistent with
both a compact and an optically thin synchrotron jet
($\alpha=-0.36\pm0.34$). If the source is in a HIMS and the radio
spectrum is optically thin, e.g., the outer part of the jet is already
decoupled from the system and the assumption of a radio compact jet is
not valid anymore, the jet model in the present form cannot be used to
fit the spectrum. On the other hard, if the source is in a hard state
and the radio jet is still optically thick, the almost flat
radio-to-IR spectral index ($\alpha=0.08\pm0.03$) is consistent with a
compact jet and the jet model should in principle be able to fit this
observation reasonably well, as it does for the other hard state
observations of GRO~J1655-40 and other BHs. The SED of March 10 seems,
however, to have a very high X-ray-to-radio flux ratio -- too high for
the model to find a good set of parameters to reproduce it. In this
case, it is tempting to claim that, therefore, at least part of the
hard X-ray emission, assumed in the jet model to be produced at the
base of the jet, does actually come from a physically different
emitting source.

We must note that on March 10, GRO~J1655-40 is already starting the
X-ray rise of the outburst, the disk emission is higher than that of
the other hard state spectra, and only a few days later the source is
in a thermal state. Therefore, a possible explanation for the bad fit
is that, e.g., the compact jet is already fading, while the accretion
rate, and thus the soft X-ray emission is increasing. However, we
still encounter the problem of explaining the high non-thermal hard
X-ray flux in the spectrum coming from a fading jet. Given the power
in the compact jet, the model cannot find a good set of parameters
that reproduces the hard X-ray emission observed. Therefore, {\it if}
the source is in a hard state with a compact radio jet, and if the
other assumptions in the model are not significantly incorrect, we
appear to observe an alternative source of hard X-rays, other than the
jet. However, since the March 10 observation is possibly in a
transitional state, a time-dependent jet model may be required in this
case.  Such a model is currently under development, and we plan to
re-visit the March 10 SED when it is available.

\section{Conclusions}

We have analysed five multi-wavelength observations of GRO~J1655-40
during its outburst in 2005. The unprecedented coverage from the radio
band to X-rays, and especially the new inclusion of
simultaneous optical/IR and mid-IR observations, allowed us to give
new constraints on the jet in a BH XRB. 

\begin{itemize}

\item We detect, for the first time, emission from a compact 
jet in the mid-IR, at 24~$\mu$m, with Spitzer/MIPS. 

\item We obtain a strong constraint on the spectral index of the 
compact jet, which is consistent with being flat from the radio band
to the mid-IR: $\alpha=0.07\pm0.04$ for the September 29 observation
and, possibly (see caveats in \S~3.1), $\alpha=0.08\pm0.03$ for the
March 10 observation.

\item Using the broadband SEDs in the hard state, we tested the  
jet scenario. We find good fits for two out of the three hard state
observations. The physical parameters of the jet are overall similar
to those previously found in other two BH XRBs, Cyg X-1 and
GX339-4. The most notable exception is the jet power, which seems to
be a factor of at least 3-5 times higher, depending on the distance,
in GRO~J1655-40, for comparable disk luminosities. We also note that
the radio-to-IR power-law spectrum observed seems to be somewhat
flatter than the model predicts.

\item The jet model does not give a good fit for the  observation on March 10, 
which can be either in a hard state or in a hard intermediate state,
depending on X-ray state definitions we adopt, and has an
unconstrained radio spectral index consistent with either a compact or
a detached optically thin radio jet. In the case of a hard state
observation, with a compact jet emitting an almost flat spectrum from
the radio band to the mid-IR, this model-data discrepancy might be
explained by the presence of an alternative source of hard X-rays,
other than the jet.

\end{itemize} 

The results presented in this work show how the wide energy range
covered by our multiwavelength campaign, with a particular key role
played by the mid-IR observations, can give strong constraints on the
radiative components of XRB systems.  The SEDs will be further used to
test improved jet models (with e.g., the inclusion of an irradiated
disk, of time-dependent parameters), as well as other disk and jet
models based on different scenarios (e.g., Yuan et al. 2005), as soon
as they can be statistically tested over the whole broadband energy
spectrum.

\acknowledgments 
 
This work is based on observations made with the Spitzer Space
Telescope, which is operated by the Jet Propulsion Laboratory,
California Institute of Technology under a contract with NASA. Support
for this work was provided by NASA through an award issued by
JPL/Caltech (\#1269434 and \#1271245). The National Radio Astronomy
Observatory is a facility of the National Science Foundation operated
under cooperative agreement by Associated Universities, Inc.
The collection and analysis of the SMARTS
data reported here was supported by grant AST-0407063 from 
the US National Science Foundation to CDB.
JAT acknowledges partial support from RXTE
cycle 9 and 10 grants NNG05GB22G and NNG06GA81G.

\begin{figure} 
\begin{center} 
\includegraphics[angle=0,scale=0.5]{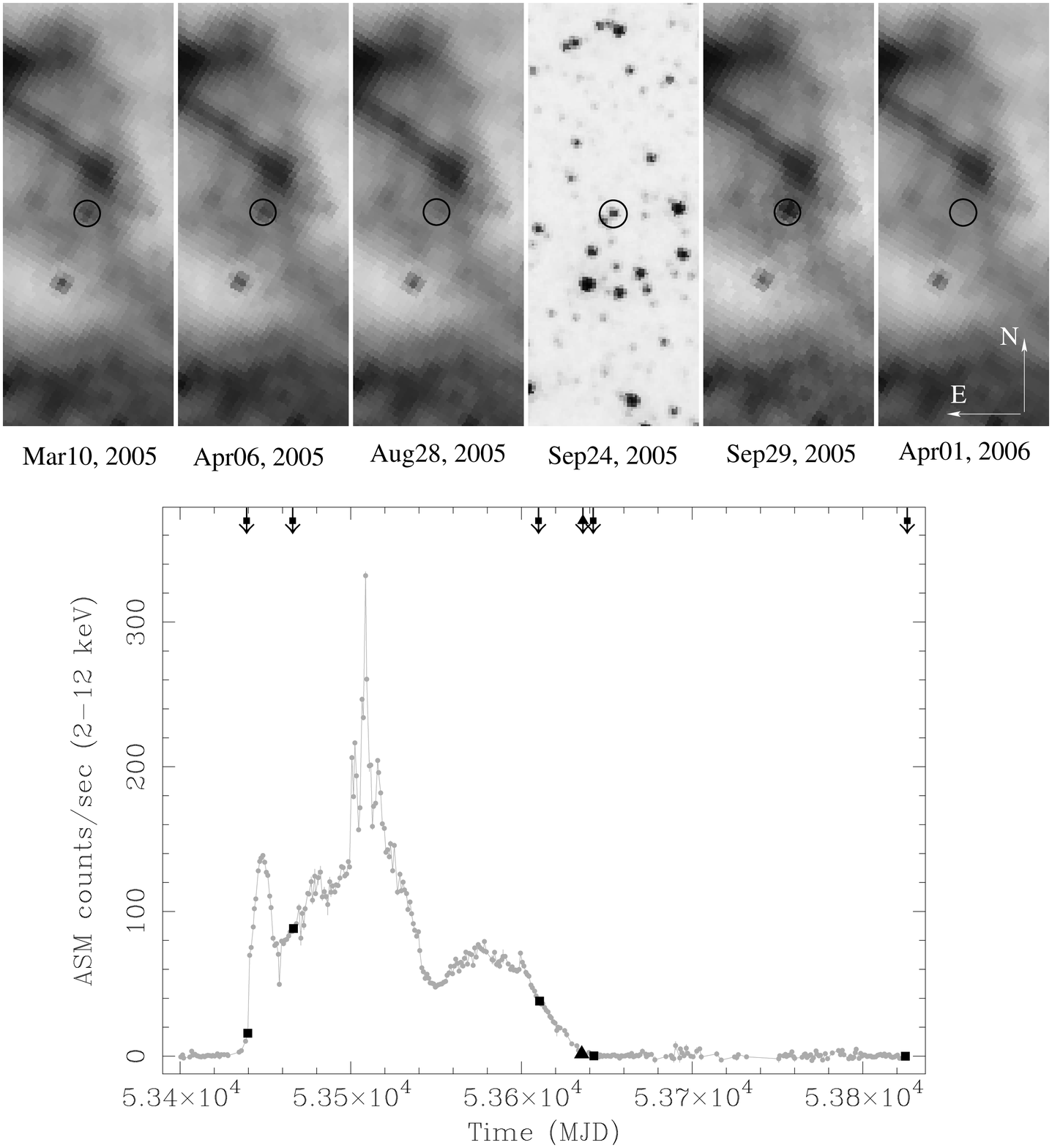} 
\caption{{\it Top panels}: Spitzer images of GRO~J1655-40. All the images are taken with MIPS at $24~\mu$m, except the one on September 24 which was taken with  IRAC (shown is the image at 4.5~$\mu$m). The optical position of GRO~J1655-40 is centered in the circle. {\it Lower panel}: RXTE/ASM light curve of the outburst. The squares represent the MIPS observations, the triangle shows the IRAC observation.} 
\label{fig:licu} 
\end{center} 
\end{figure} 

\begin{figure} 
\begin{center} 
\includegraphics[angle=0,scale=0.7]{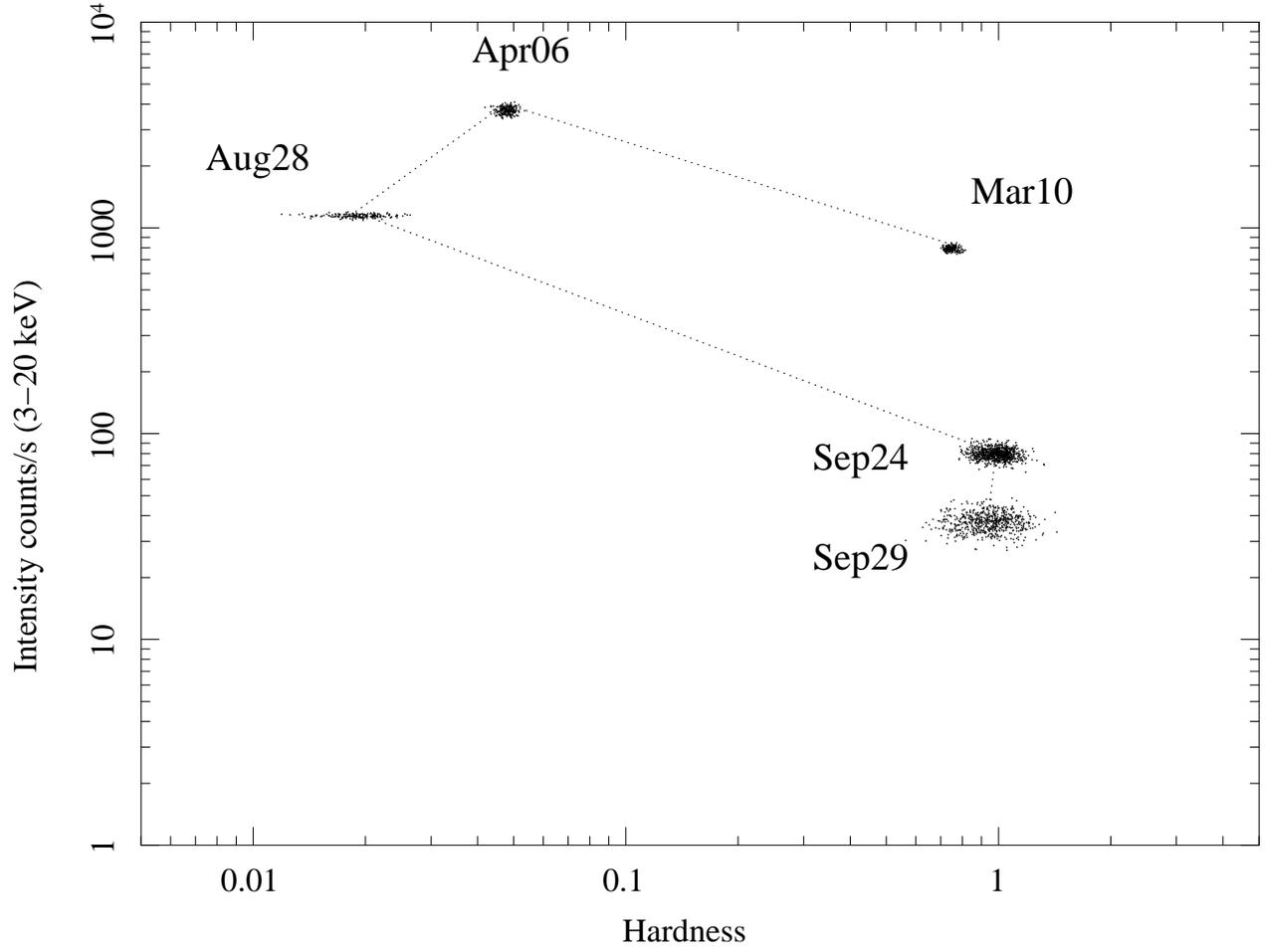} 
\caption{Hardness-Intensity Diagram (HID) of the five RXTE pointed observations,   
simultaneous with Spitzer.} 
\label{fig:hid} 
\end{center} 
\end{figure} 

\begin{figure} 
\begin{center} 
\includegraphics[angle=0,scale=0.95]{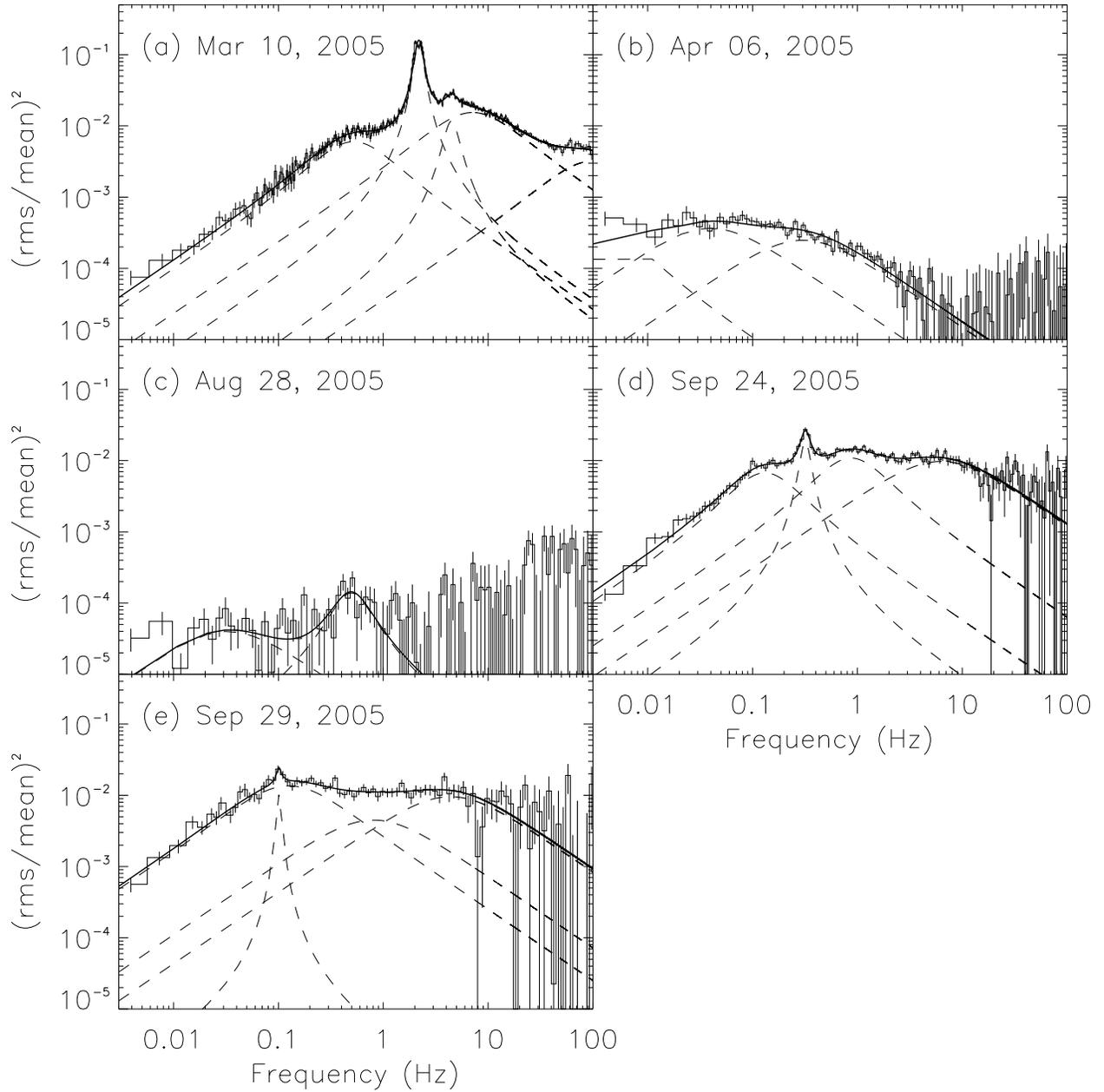} 
\caption{X-ray power density spectra of the five RXTE/PCA observations simultaneous with Spitzer. The dashed lines indicate individual Lorentzian components whereas the solid lines show the total fit.} 
\label{default} 
\end{center} 
\end{figure} 

\begin{figure} 
\begin{center} 
\includegraphics[angle=0,scale=0.35]{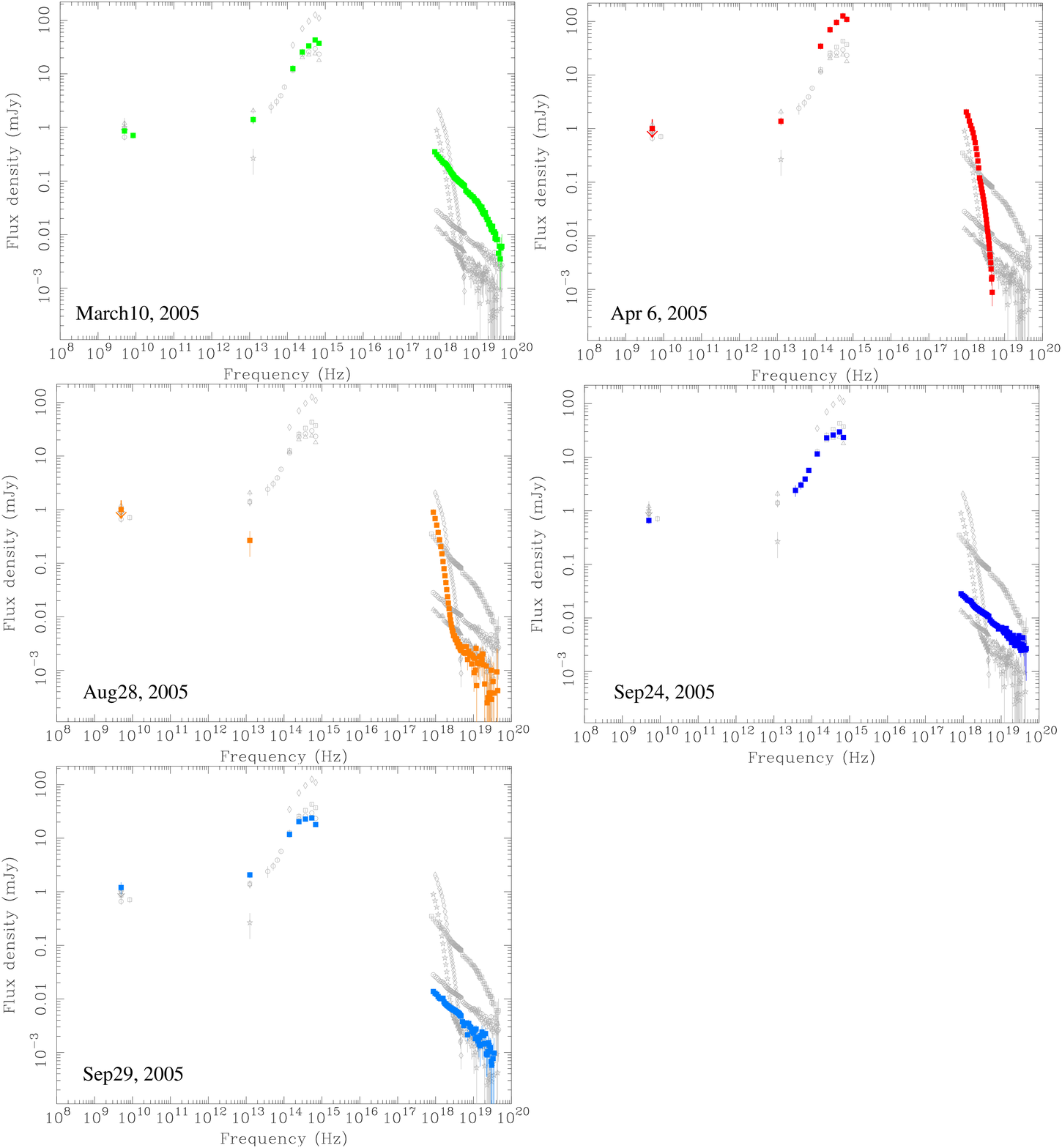} 
\caption{Broadband energy spectra of GRO~J1655-40. The gray open markers show all the five spectra  
with Spitzer coverage during the outburst, and the filled markers
highlight the observation on the date indicated in the bottom-left
corner of each panel. }
\label{default} 
\end{center} 
\end{figure} 

\begin{figure}
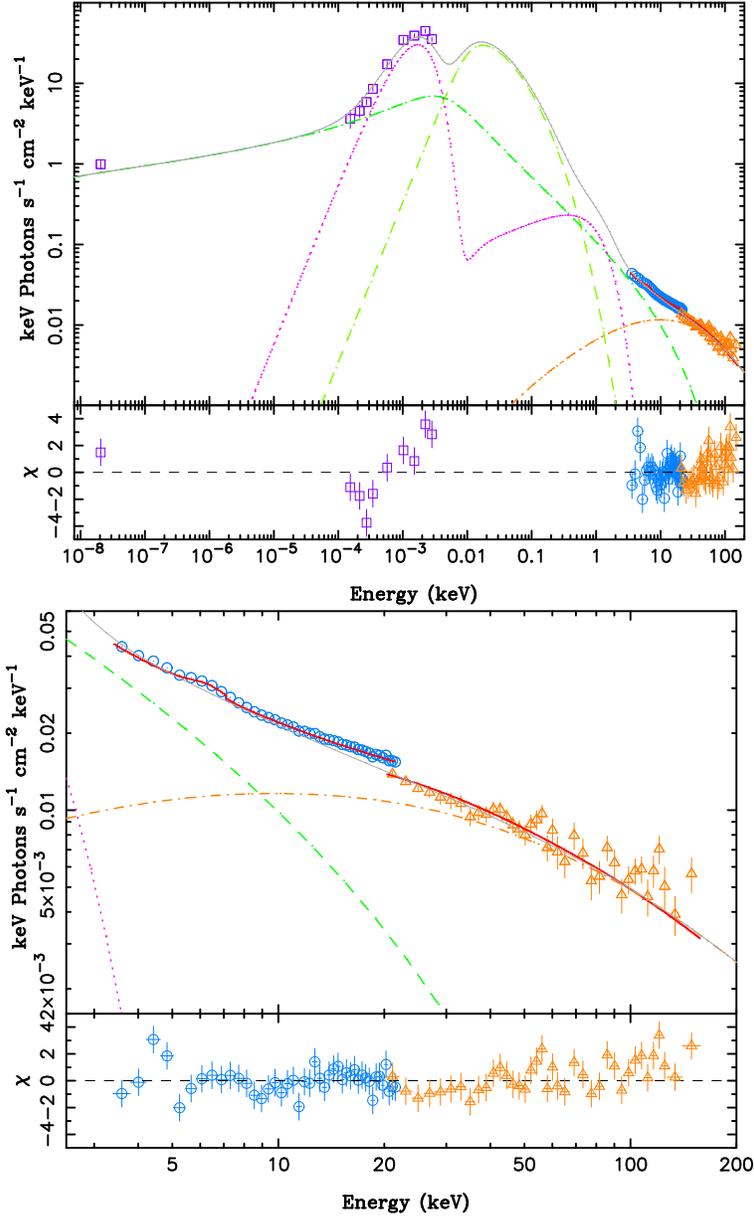
 
\begin{center} 
\includegraphics[angle=0,scale=0.5]{f5a.ps}\\ 
\includegraphics[angle=0,scale=0.5]{f5b.ps}\\ 
\caption{Jet model fits with residuals of the radio-to-X-ray (upper panel) and  X-ray  spectrum  
(lower panel) of the September 24 observation of GRO~J1655-40, with
the jet inclination angle fixed to $75^{\circ}$. The light green
dashed line is the pre-shock synchrotron component, the darker green dash-dotted
line is the post-shock synchrotron component, the orange dash-dotted line
represents the SSC plus the disk external Compton component,
the purple dotted line is the multi-temperature disk black body plus a
black body representing the companion star in the binary system. The
solid red line is the total model. Note that the model components in
this representation are not absorbed and are not convolved with the
response matrices. } 
\label{sep24fit} 
\end{center} 
\end{figure} 

\begin{figure} 
\begin{center} 
\includegraphics[angle=0,scale=0.6]{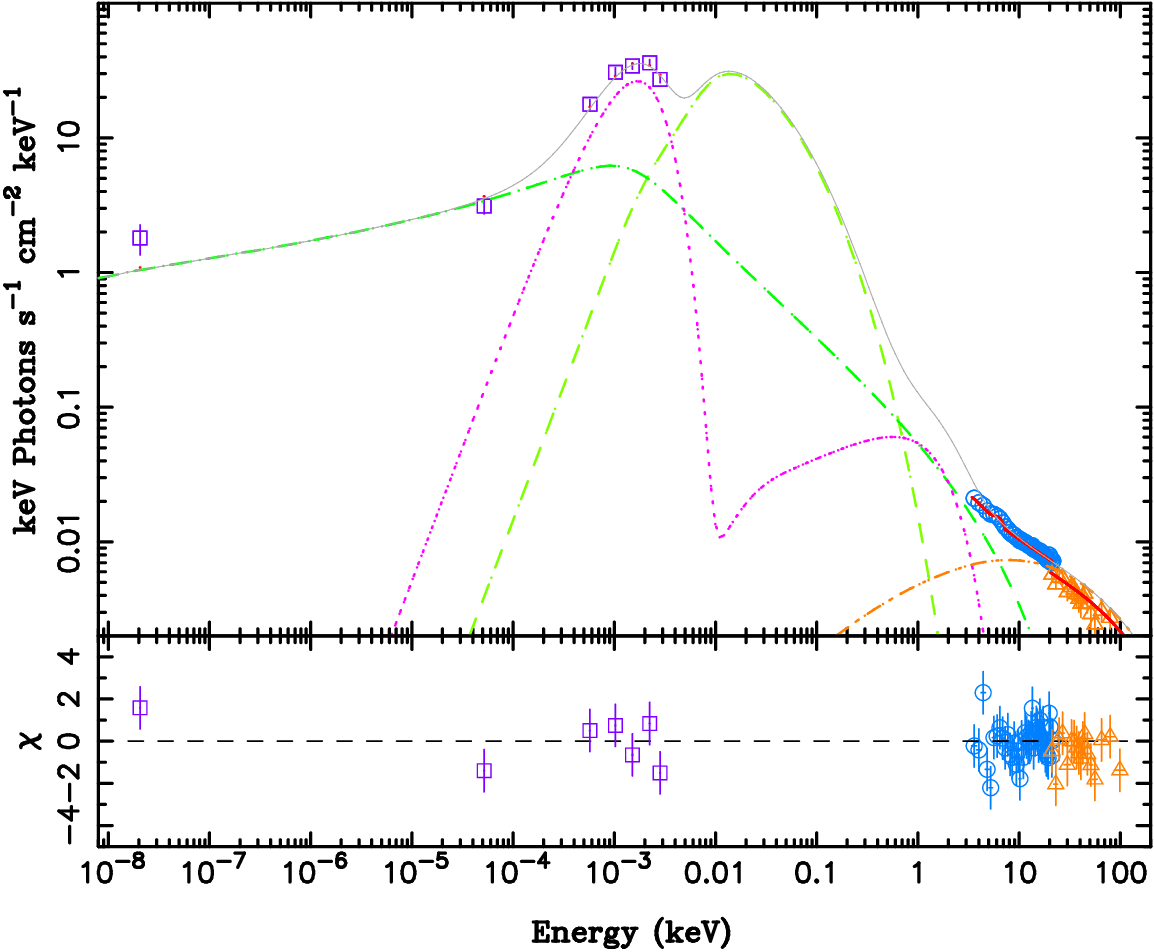}\\ 
\includegraphics[angle=0,scale=0.6]{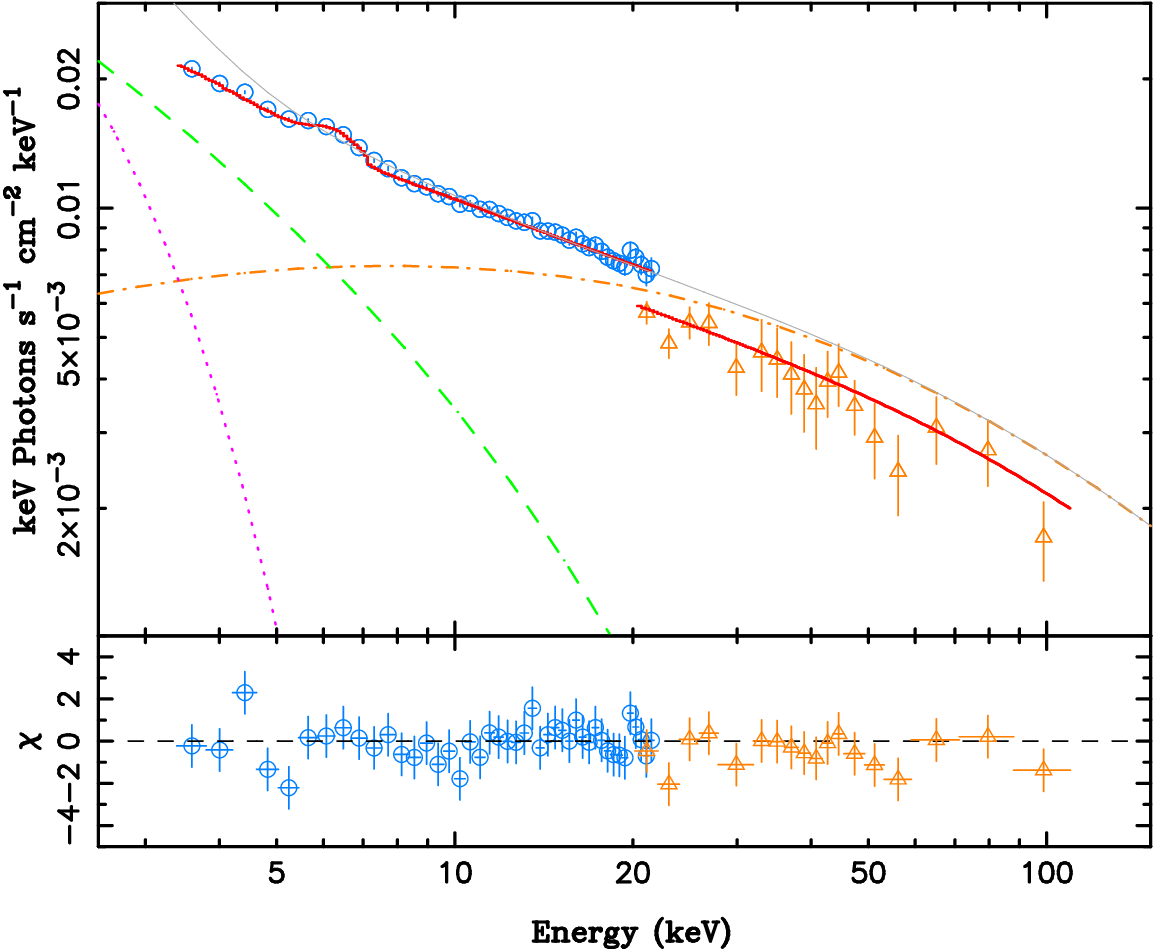}\\ 
\caption{Jet model fits with residuals of the radio-to-X-ray (upper panel) and X-ray (lower panel)  
spectrum of the September 29 observation of GRO~J1655-40, with the jet 
inclination angle fixed to $75^{\circ}$. Model components as in Fig.~\ref{sep24fit}.} 
\label{sep29fit} 
\end{center} 
\end{figure} 
 
\begin{figure} 
\begin{center} 
\includegraphics[angle=0,scale=0.6]{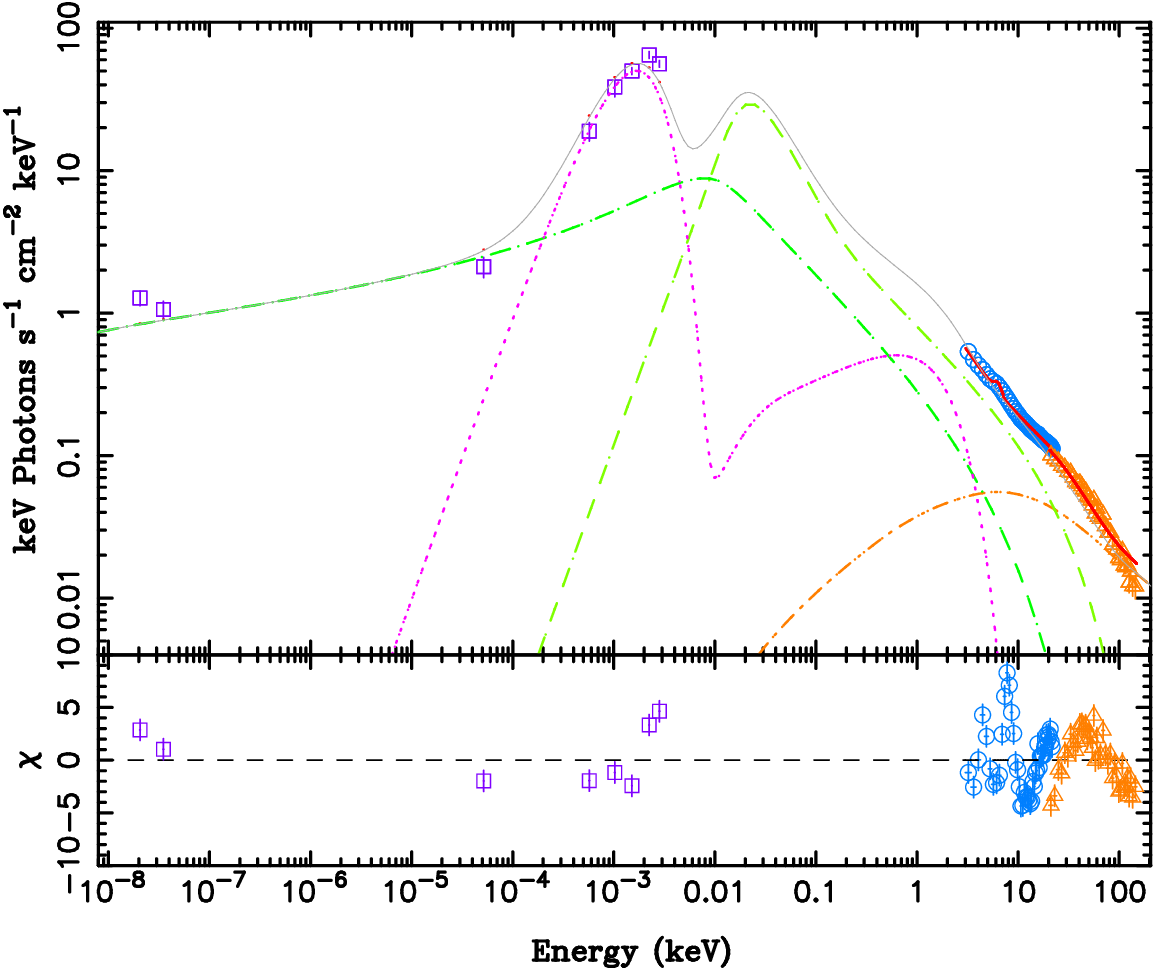}\\ 
\includegraphics[angle=0,scale=0.6]{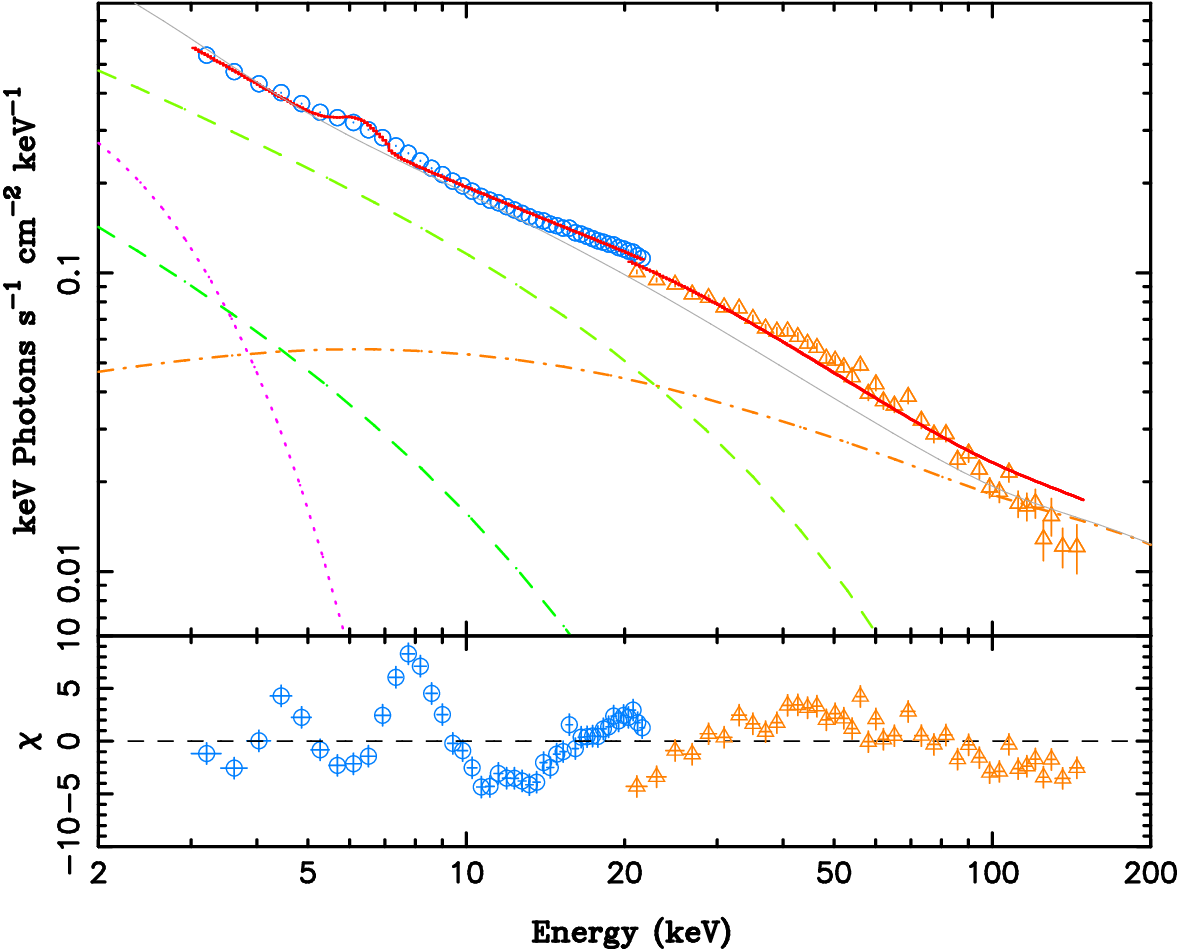}\\ 
\caption{Jet model fits with residuals of the radio-to-X-ray (upper panel) and X-ray (lower panel)  
spectrum of the March 10 observation of GRO~J1655-40, with the jet 
inclination angle fixed to $75^{\circ}$.  Model components as in Fig.~\ref{sep24fit}.} 
\label{sep29fit} 
\end{center} 
\end{figure} 

 
\begin{table*}

\caption{Logs of the pointed RXTE and Spitzer simultaneous observations. Spitzer MIPS observations are at 24~$\mu$m, Spitzer IRAC are at 3.6, 4.5, 5.8 and 8~$\mu$m. For the  Spitzer observations, we show the exposure time per pixel per Basic Calibrated Data (BCD), times the number of BCD frames used to create the  mosaics.} 
 
\vspace{0.5cm} 
\begin{tabular}{l l l l} 
\tableline 
 
\hline 
\multicolumn{4}{c}{RXTE} \\
\hline 
Instrument & Start Obs. & End Obs. & ObsID\\ 
PCA/HEXTE& 2005-03-10UT17:45:36& 2005-03-10UT18:46:40 & 90704-04-01-00\\
PCA/HEXTE& 2005-04-06UT00:47:28& 2005-04-06UT01:53:36 & 91702-01-24-00\\ 
PCA/HEXTE& 2005-04-06UT02:16:16& 2005-04-06UT02:57:36 & 91702-01-24-02\\ 
PCA/HEXTE& 2005-04-06UT10:27:12& 2005-04-06UT10:50:40 & 91702-01-24-03\\ 
PCA/HEXTE& 2005-08-28UT22:59:12& 2005-08-29UT00:33:36 & 91702-01-47-10\\
PCA/HEXTE& 2005-08-28UT12:08:48& 2005-08-28UT13:12:48 & 91702-01-47-12\\ 
PCA/HEXTE& 2005-08-28UT21:16:16& 2005-08-28UT22:05:36 & 91702-01-46-12\\ 
PCA/HEXTE& 2005-09-24UT05:42:24& 2005-09-24UT11:23:44 & 91704-01-01-01\\
PCA/HEXTE& 2005-09-29UT06:47:12& 2005-09-29UT10:53:36 & 91702-01-87-02\\
\hline 
\multicolumn{4}{c}{Spitzer} \\ 
\hline 
Instrument & Start Obs. & Exp. Time$\times$BCDs&  ObsID \\ 
MIPS & 2005-03-10UT18:13:34&  9.96 s $\times70$& 10525184\\ 
MIPS& 2005-04-06UT14:17:07& 9.96 s $\times70$& 10525440\\ 
MIPS& 2005-08-28UT11:43:36& 9.96 s $\times70$& 10525696\\ 
IRAC& 2005-09-23UT10:47:11& 1.2 s $\times59$& 14508800\\ 
MIPS& 2005-09-29UT02:43:03& 9.96 s $\times70$& 14509312\\ 
MIPS& 2006-04-01UT21:55:03& 9.96 s $\times70$& 10524928\\ 

\tableline

\end{tabular} 

\end{table*} 
 
 
\begin{table*}

\caption{ Date, wavelength of observation, X-ray state and flux densities of the Spitzer MIPS observations at 24~$\mu$m and  Spitzer IRAC at 3.6, 4.5, 5.8 and 8~$\mu$m. Errors are 1$\sigma$ rms.}
 
\vspace{0.5cm} 

\begin{tabular}{l l l l} 
\tableline 
Date & Wavelength & X-ray state & Flux density (mJy) \\
\tableline
2005 March 10& 24 $\mu$m & hard/HIMS & $1.40\pm0.23$\\ 
2005 April 6& 24 $\mu$m & thermal & $1.37\pm0.22$\\ 
2005 August 28& 24 $\mu$m & thermal & $0.26\pm0.13$\\ 
2005 September 23 & 3.6 $\mu$m&hard &$5.67\pm0.39$\\
                                 & 4.5 $\mu$m& &$3.90\pm0.29$\\
                                 & 5.8 $\mu$m& &$3.02\pm0.52$\\
                                 & 8 $\mu$m   & &$2.40\pm0.57$\\ 
2005 September 29& 24 $\mu$m & hard & $2.07\pm0.27$\\ 
2006 April 1 & 24 $\mu$m & quiescence & $<0.54$\\ 
\tableline  
 
\end{tabular} 
\end{table*} 
 
 
\begin{table*}

\caption{ Start time and apparent (non de-reddened) magnitudes of the B, V, I, J and K band observations of GRO~J1655-40 with SMARTS. For each observing run, the exposure time is 6 minutes for the B, J and K bands, 4 minutes for the V band and 2 minutes for the I band. See also \S~2.2.}
 
\vspace{0.5cm} 

\begin{tabular}{l l l l l l} 
\tableline 
Start Time                                            & B mag& V mag& I mag& J mag& K mag\\
\tableline
2005-03-09UT07:25:28&	17.80&	16.17&	14.03&	13.05 &  12.21\\
2005-03-10UT08:09:14&	17.62&	16.04&	13.99&	13.06 &  12.23\\
2005-04-05UT06:52:23&	16.28&	14.77&	12.73&	11.91&   11.07\\
2005-04-06UT06:18:04&	16.45&	14.87&	12.84&	11.97&   11.14\\
2005-04-07UT06:14:46&	16.30&	14.76&	12.71&	11.87&   11.04\\
2005-09-22UT00:26:25&	18.12&	16.45&	14.26&	13.18 &   12.33\\
2005-09-28UT00:30:41&	18.54&	16.63&	14.38&	13.47 &  12.47\\
2005-09-29UT00:39:09&	18.41&	16.68&	14.41&	13.31&   12.30\\
2005-09-30UT23:56:40&	18.16&	16.48&	14.25&	13.17 &  12.27\\
\tableline  
 
\end{tabular} 

\end{table*} 
 

\begin{table*} 
\caption{Best-fit parameters of the RXTE energy spectra of GRO~J1655-40. Errors are 68\% confidence for one interesting parameter.}

\begin{tabular}{l l l l l l} 
\tableline 
\tableline  Parameter         & Mar10 & Apr6 & Aug28 & Sep24 & Sep29 \\ 
\tableline

\hline 
\multicolumn{6}{c}{Disk Blackbody} \\ 
\hline 
 
kT$_{BB}$ (keV) & $0.5~fixed$ &$1.25^{+0.01}_{-0.01}$ & $0.87^{+0.01}_{-0.01}$&  $0.5~fixed$ & $0.5~fixed$\\ 
 
N$_{BB}$                            & $1969^{+139}_{-140}$ &$718^{+26}_{-29}$  & $1904^{+15}_{-40}$&  $115^{+29}_{-29}$ & $106^{+18}_{-17}$  \\

\hline 
\multicolumn{6}{c}{Gaussian} \\ 
\hline 
E$_{Fe}$ (keV)                           &  $--$ &  $6.24^{+0.07}_{-0.10}$& $--$          &$--$           & $--$\\ 
 
$\sigma_{Fe}$ (keV)                      & $--$  & $0.1~fixed$ & $--$ &$--$  & $--$\\ 
 
N$_{Fe}$   $(\times 10^{-3}$ phot. cm$^{-2}$ s$^{-1})$      & $--$  & $9.02^{+0.01}_{-0.01}$ & $--$ &$--$  & $--$\\ 
 
\hline 
\multicolumn{6}{c}{Power Law} \\ 
\hline

$\Gamma_{PL}$                         & $1.72^{+0.01}_{-0.01}$  & $2.93^{+0.08}_{-0.07}$     & $2.28^{+0.07}_{-0.05}$& $1.57^{+0.01}_{-0.01}$   & $1.57^{+0.01}_{-0.02}$ \\ 
 
N$_{PL}$  (10$^{-12}$ erg cm$^{-2}$ s$^{-1}$)    & $4296^{+13}_{-13}$  & $5025^{+700}_{-636}$    & $498^{+42}_{-39}$& $462^{+2}_{-2}$  & $ 205^{+1}_{-1}$   \\

\hline 
\multicolumn{6}{c}{High-Energy Cutoff} \\ 
\hline 
 
E$_{cut}$  (keV)               & $33^{+3}_{-3}$           & $11.9^{+0.2}_{-0.2}$ & $--$ &$--$   & $--$\\ 
 
E$_{fold}$  (keV)              & $160^{+12}_{-11}$      & $4.6^{+0.3}_{-0.9}$           & $--$  &$--$  & $--$\\

\hline 
\multicolumn{6}{c}{Smedge} \\ 
\hline 
 
E$_{edge}$  (keV)                     & $7.44^{+0.06}_{-0.06}$ &$7.86^{+0.09}_{-0.06}$   & $8.64^{+0.08}_{-0.07}$& $6.8^{+0.1}_{-0.1}$  & $6.7^{+0.2}_{-0.2}$ \\ 
 
Max$\tau$                  & $1.81^{+0.05}_{-0.05}$ &$1.0^{+0.3}_{-0.3}$  & $5.0^{+0.4}_{-0.4}$& $1.46^{+0.08}_{-0.08}$ & $0.6^{+1.2}_{-0.3}$ \\ 
 
Width$_{edge}$  (keV)             & $>10.6$                         &$1.4^{+0.6}_{-0.5}$  & $>8.7$& $>7.5$ & $5.6^{+11.7}_{-2.8}$ \\ 
\hline 
\multicolumn{6}{c}{Fit Statistics} \\ 
\hline 
 
$\chi^{2}_{\nu}$ (d.o.f.)                        & $1.09~(85)$              & $1.53~(38)$              & $ 1.26~(85)$&  $0.64~(86)$              &  $0.73~(86)$\\

\tableline  
 
\end{tabular} 
\end{table*} 
 
 
 
 {\rotate
\begin{table} 
\caption{Jet model best-fit parameters of the broadband energy spectra of GRO~J1655-40 on 2005 September 24 and September 29, with a fixed distance of 1.7~kpc. One fit of the September 29 observation with a distance of 3.2 kpc is also shown. The best-fit parameters of two observations of GX~339-4 and Cyg~X-1 from Markoff, Nowak \& Wilms (2005) are  reported for comparison. Errors are 90\% confidence level. $N_{j}$ is the jet normalization in Eddington luminosity units and is of the same order of the jet power, $r_{0}$ is the nozzle radius in gravitational radii units, $T_{e}$ is the temperature of the pre-shocked electrons, $p$ is the spectral index of the post-shock electron power-law distribution, $k$ is the equipartition factor in units of the ratio between magnetic and electron energy density, $pl_{f}$ is the fraction of accelerated electrons, $z_{acc}$ is the location of the acceleration region in gravitational radii units, and  $h_{0}$ is the radius-to-height ratio of the nozzle region. The $u_{acc}/c$ parameter is the shock speed relative to the bulck plasma flow, $f_{sc}$ is the ratio of the scattering mean free path to the gyroradius, $L_{disk}$ is the disk luminosity in units of Eddigton luminosity, $T_{disk}$ is the inner disk temperature, $A_{line}$ , $E_{line}$ and  $\sigma_{line}$ are, respectively,  the normalization, the peak energy and the width of the Gaussian emission line, and $\Omega/2\pi$ is the reflection fraction.}

{\tiny
\begin{tabular}{l l l l l l l l l l} 
\tableline 
\tableline   
Date   &  $N_{j}$                   &  $r_{0}$            & $T_{e}$       & $p$ &$k$ &$pl_{f}$ &$z_{acc}$ & $h_{0}$&\\ 
(jet incl./dist.)  &  ($10^{-3}L_{Edd}$) &  ($GM/c^{2}$) & ($10^{10}K$) &  & & & & &\\ 
\tableline 
 \hline 
\multicolumn{9}{c}{GRO J1655-40} \\ 
\hline 
                 Sep24 ($75^\circ$/1.7) &  2.91$_{-0.04}^{+0.01}$    &  3.48$_{-0.01}^{+0.01}$    &  4.82$_{-0.01}^{+0.02}$  & 2.50$_{-0.01}^{+0.01}$  &   4.17$_{-0.01}^{+0.01}$ & 0.75 (fixed) & 7 (fixed) &  1.4 (fixed) &\\ 
                 Sep29 ($75^\circ$/1.7) &   2.94$_{-0.02}^{+0.01}$ &   4.44$_{-0.02}^{+0.03}$  &  4.87$_{-0.03}^{+0.02}$  & 2.37 $_{-0.01}^{+0.01}$  & 5.41$_{-0.02}^{+0.04}$ & 0.75 (fixed) &20 (fixed)& 1.2 (fixed) &\\ 
                 Sep29 (free/1.7) &   2.23$_{-0.01}^{+0.01}$   &   2.55$_{-0.11}^{+0.24}$   &   4.23 $_{-0.04}^{+0.01}$  & 2.40$_{-0.01}^{+0.01}$ & 1.15$_{-0.05}^{+0.01}$& 0.75 (fixed)& 20 (fixed) & 1.2 (fixed) &\\ 
Sep29 (free/3.2)    &   3.68$_{-0.01}^{+0.01}$  &   2.08$_{-0.08}^{+0.46}$   &   3.10$_{-0.01}^{+0.01}$ & 2.55$_{-0.01}^{+0.01}$ & 1.29$_{-0.02}^{+0.01}$ & 0.75 (fixed)& 10 (fixed) & 1.5 (fixed) &\\ 
 \hline 
\multicolumn{9}{c}{GX 339-4} \\ 
\hline 
Apr2, 1999 &  0.64$^{+0.02}_{-0.03}$ & 9.6$^{+0.5}_{-0.1}$   & 5.23$^{+0.13}_{-0.12}$  & 2.39$^{+0.01}_{-0.01}$  & 1.12$^{+0.01}_{-0.01}$  & 0.74$^{+0.05}_{-0.01}$  & 302  & 1.41$^{+0.01}_{-0.01}$ & \\
 \hline 
\multicolumn{9}{c}{Cyg X-1} \\ 
\hline 
Feb23, 2003     &  0.74$^{+0.01}_{-0.01}$ & 4.4$^{+0.2}_{-0.1}$  & 3.28$^{+0.01}_{-0.01}$  & 2.61$^{+0.01}_{-0.01}$  & 1.77$^{+0.04}_{-0.01}$  & 0.73$^{+0.08}_{-0.02}$  & 9  & 1.18$^{+0.01}_{-0.01}$ &\\

\tableline 
\tableline   
 Date  &  $u_{acc}/c$ &  $f_{sc}$ & $L_{disk}$              & $T_{disk}$ &$A_{line}$   &$E_{line}$ & $\sigma_{line}$ &$\Omega/2\pi$& $\chi^{2}_{\nu} (dof)$\\ 
 (jet incl./dist)          &                     &                & ($10^{-3}L_{Edd}$) & (keV)    &($10^{-2}$) & (keV)      & (keV)                  &                          & \\ 
\tableline 
 \hline 
\multicolumn{10}{c}{GRO J1655-40} \\
\hline 
 Sep24 ($75^\circ$/1.7) &  0.6 (fixed) &  945$_{-51}^{+8}$  &0.36$_{-0.13}^{+0.39}$  &   0.39 $_{-0.25}^{+0.12}$   & 0.04$_{-0.01}^{+0.01}$& 6.2 (fixed) &  0.6 (fixed) & $0.36_{-0.02}^{+0.13}$ & 1.72 (81)\\ 
 Sep29 ($75^\circ$/1.7) &  0.6 (fixed)  &  2450 $_{-77}^{+58}$   & 0.15$_{-0.01}^{+0.01}$  & 0.62 $_{-0.02}^{+0.04}$    & 0.02 $_{-0.01}^{+0.01}$ &  6.2 (fixed)&  0.5 (fixed)& $<0.01$ & 0.90 (56)\\     
 Sep29 (free/1.7)    & 0.6 (fixed)   & 420$_{-8}^{+1}$   &   0.07$_{-0.01}^{+0.01}$ &  0.59 $_{-0.02}^{+0.01}$   & 0.02 $_{-0.01}^{+0.01}$ & 6.2 (fixed) & 0.65 (fixed)  & 0.37$_{-0.02}^{+0.03}$ & 0.94  (55)\\ 
Sep29 (free/3.2)   & 0.6 (fixed)   & 631$_{-13}^{+336}$   &   0.45$_{-0.03}^{+0.03}$ &  0.54$_{-0.02}^{+0.01}$  & 0.01$_{-0.01}^{+0.01}$ & 6.1$_{-0.1}^{+0.2}$ & 0.65 (fixed)  & 0.61$_{-0.07}^{+0.03}$ & 1.20  (53)\\

 \hline 
\multicolumn{10}{c}{GX 339-4} \\ 
\hline 
Apr2, 1999 &  0.32$^{+0.05}_{-0.01}$ & 1100$^{+200}_{-800}$   & 0.33$^{+0.01}_{-0.01}$  & 1.53$^{+0.12}_{-0.10}$  & 0.09$^{+0.03}_{-0.03}$  & 6.4$^{+0.1}_{-0.1}$  & 0.7$^{+0.1}_{-0.1}$  & $<0.06$  & 1.76 (87)\\
 \hline 
\multicolumn{10}{c}{Cyg X-1} \\ 
\hline 
Feb23, 2003     &  0.35$^{+0.01}_{-0.01}$ & 790$^{+10}_{-10}$  & 0.8$^{+0.1}_{-0.1}$  & 0.98$^{+0.11}_{-0.09}$  & 2.3$^{+0.2}_{-0.5}$  & 6.0$^{+0.1}_{-0.1}$  & 0.9$^{+0.1}_{-0.1}$  & $<0.01$ & 1.17 (177)\\
\tableline  

\end{tabular} 
} 
\end{table} 
 }


\begin{thebibliography}{} 
 
\bibitem[]{} 
Belloni T., Psaltis D., van der Klis M., 2002, ApJ, 572, 392 
\bibitem[]{}
Beloborodov A.M., 1999, ApJ, 510, L123
\bibitem[]{} 
Buxton M., Bailyn C.D., Maitra D., 2005, ATel, 408
\bibitem[]{} 
Buxton M., Bailyn C.D., 2005, ATel, 608
\bibitem[]{} 
Buxton M., Bailyn C.D., Maitra D., 2005, ATel, 418 
\bibitem[]{} 
Buxton M., Bailyn C.D., 2004, ApJ, 615, 880
\bibitem[]{}  
Brocksopp K., et al., 2005, ATel., 612 
\bibitem[]{} 
Cardelli J.A., Clayton, G.C., Mathis, J.S., 1989, ApJ, 345, 245 
\bibitem[]{} 
Corbel S., Fender R. P., Tomsick J. A., Tzioumis A. K., Tingay S., 2004, ApJ, 617, 1272 
\bibitem[]{} 
Cunningham C., 1976, ApJ, 208, 534
\bibitem[]{}
Dhawan V., Mirabel I.F., Rodr\'\i{}guez L.F., 2000, ApJ, 543, 373
\bibitem[]{}
Depoy et al. 2003, SPIE, 4841, 827
\bibitem[]{} 
Fender R.P., 2006, Compact Stellar X-Ray Sources, eds. W.H.G. Lewin
and M. van der Klis, Cambridge University Press, 381
\bibitem[]{}
Fender R. P., Belloni T. M., Gallo E., 2004, MNRAS, 355, 1105
\bibitem[]{} 
Fender, R.P., 2001, MNRAS, 322, 31 
\bibitem[]{}
Fender R.P. et al., 1999, ApJ, 519, L165
\bibitem[]{} 
Greene, J., Bailyn, C.D., Orosz, J.A., 2001, ApJ, 554, 1297 
\bibitem[]{}
Hynes R. I., Haswell C. A., Chaty S., Shrader C. R., Cui W., 2002, MNRAS, 331, 169
\bibitem[]{}
Heinz S., 2004, MNRAS, 355, 835
\bibitem[]{} 
Homan, J., 2005, ATel, 440 
\bibitem[]{} 
Homan, J., Miller, J.M., Wijnands, R., Lewin, W.H.G., 2005a, ATel, 487 
\bibitem[]{} 
Homan, J., Miller, J.M., Wijnands, R., Lewin, W.H.G., 2005b, ATel, 607 
\bibitem[]{} 
Homan, J., 2005c, ATel, 440 
\bibitem[]{} 
Houck J.C., Denicola L.A., 2000, ASPC, 216, 591 
\bibitem[]{} 
Hjellming, R.M., Rupen, M.P., 1995, Nature, 375, 464  
\bibitem[]{} 
Kalemci, E. et al., 2005, ApJ, 622, 508  
\bibitem[]{} 
Landolt 1992, AJ, 104, 340
\bibitem[]{} 
Maccarone T.J., 2002, MNRAS, 336, 1371 
\bibitem[]{} 
Maccarone T.J, 2005, MNRAS, 360, L68
\bibitem[]{} 
Makovoz, D., \& Marleau, F., 2005, PASP, 117, 1113 
\bibitem[]{}
Markoff, S.; Nowak, M.; Falcke, H.; Maccarone, T.; Fender, R., 2004,
Nuclear Physics B Proceedings Supplements, 132, 129
\bibitem[]{} 
Markoff, S., Nowak, M.A., 2004, ApJ, 609, 972  
\bibitem[]{} 
Markoff, S., Nowak, M.A., Wilms, J., 2005, ApJ, 635, 1216 
\bibitem[]{} 
Markwardt, C.B., Swank, J.H., 2005, ATel, 414 
\bibitem[]{} 
McClintock, J.E., \& Remillard, R.A., 2006, Compact Stellar X-Ray Sources, 
eds. W.H.G. Lewin and M. van der Klis, Cambridge University Press, 157, 
astro-ph/0306213 
\bibitem[]{} 
Meier, D.L., 2001, ApJ, 538, L9 
\bibitem[]{} 
Miyamoto, S., Kitamoto, S., 1989, Nature, 342, 773 
\bibitem[]{} 
Nowak M.A., J. Wilms., Heinz S., Pooley G., Pottschmidt K., Corbel S., 2005, ApJ, 626, 1006 
\bibitem[]{}
Nowak M.A., Wilms J., Vaughan B.A., Dove J. B., Begelman M.C., 1999, ApJ, 522, 460
\bibitem[]{}
Nowak M.A., 2000, MNRAS, 318, 361
\bibitem[]{}
Pottschmidt K., Wilms J., Nowak M. A., Pooley G. G., Gleissner T., Heindl W. A., Smith D. M., Remillard R., Staubert R., 2003, A\&A, 407, 1039
\bibitem[]{} 
Pottschmidt, K., 2002, PhD Thesis, Univ. of T\"ubingen 
\bibitem[]{} 
Reach W.T., et al., 2005, PASP, 117, 978 
\bibitem[]{} 
Remillard R.A., McClintock J.E., 2006, ARA\&A, 44, 49 
\bibitem[]{} 
Rieke G.H., Lebofsky M.J.,  1985, ApJ, 288, 618 
\bibitem[]{} 
Rupen, M.P., Dhawan, V., Mioduszewski, A.J., 2005a, ATel, 419 
\bibitem[]{} 
Rupen, M.P., Dhawan, V., Mioduszewski, A.J., 2005b, ATel, 441 
\bibitem[]{} 
Rupen, M.P., Dhawan, V., Mioduszewski, A.J., 2005c, ATel, 489 
\bibitem[]{}
Russell D. M., Fender R. P., Hynes R. I., Brocksopp C., Homan J., Jonker P. G., Buxton M. M., 2006, MNRAS, 371, 1334
\bibitem[]{} 
Shaposhnikov N., Swank J., Shrader C. R., Rupen M., Beckmann V.,
Markwardt C. B., Smith D. A., 2007, ApJ, 655, 434
\bibitem[]{} 
Tingay S.J., et al., 1995, Nature, 374, 141
\bibitem[]{} 
Tomsick J.A., Kaaret P., Kroeger R.A., Remillard R.A, 1999, ApJ, 512, 892
\bibitem[]{}
Tomsick J.A., Kalemci E., Kaaret P., 2004, ApJ, 601, 439
\bibitem[]{} 
Torres M.A.P., Steeghs, D., Jonker, P., Martini, P., 2005, ATel, 417 
\bibitem[]{} 
van der Klis M., 1988, in H. \"{O}gelman and E.P.J. van den Heuvel, eds, 
Proceedings of the NATO Advanced Study Institute on Timing Neutron Stars, held 
in \c{C}e\c{s}me, Izmir, Turkey, April 4--15
\bibitem[]{}
Vrtilek S. D., Raymond J. C., Garcia M. R., Verbunt F., Hasinger G., Kurster M., 1990, ApJ, 235, 162
\bibitem[]{} 
Wilms J, Nowak, M. A.; Pottschmidt, K.; Pooley, G. G.; Fritz, S., 2006, A\&A, 447, 245 
\bibitem[]{} 
Yuan F., Cui W., Narayan R., 2005, ApJ, 620, 905
\bibitem[]{}
Zdziarski, A.A., Poutanen J., Mikolajewska J., Gierlinski M., Ebisawa K., Johnson W. N., 1998, MNRAS, 301, 435
\bibitem[]{} 
Zhang W., Jahoda K., Swank J.H., Morgan E.H., Giles A.B., 1995, ApJ, 
449, 930 
\end{thebibliography}
\end{document}